\documentclass[apj]{emulateapj}

\usepackage{amsmath}
\usepackage{apjfonts}
\usepackage{gensymb}
\usepackage{ulem}
\usepackage{subfigure}
\usepackage{booktabs}
\usepackage{color}
\usepackage{graphicx}
\usepackage{epstopdf}
\usepackage[dvipsnames]{xcolor}
\usepackage{appendix}
\usepackage{xcolor, soul}

\newcommand\kms{\; {\rm km}\;{\rm s}^{-1}}

\newcommand\kpc{\;{\rm kpc}}

\newcommand\Gyr{\;{\rm Gyr}}

\newcommand{\olsi}[1]{\,\overline{\!{#1}}} 

\def\spose#1{\hbox to 0pt{#1\hss}}
\def\dt{\spose{\raise 1.0ex\hbox{\hskip2pt$\mathchar"201$}}}    

\shorttitle{Understanding the velocity distribution of the Galactic Bulge with APOGEE and Gaia}
\shortauthors{Zhou et al.}

\begin{document}



\title{Understanding the velocity distribution of the Galactic Bulge with APOGEE and Gaia}
\author{Yingying Zhou$^{1,2}$, Zhao-Yu Li$^{3,4}$, Iulia Simion$^{1}$,  Juntai Shen$^{3,4,1}$, Shude Mao$^{5,6}$, Chao Liu$^{6,2}$, Mingjie Jian$^{7}$, Jos{\'e}. G. Fern{\'a}ndez-Trincado$^{8}$}

\affil{
$^1$Key Laboratory for Research in Galaxies and Cosmology, Shanghai Astronomical Observatory, Chinese Academy of Sciences, 80 Nandan Road, Shanghai 200030, China \\
$^2$College of Astronomy and Space Sciences, University of Chinese Academy of Sciences, 19A Yuquan Road, Beijing 100049, China \\
$^3$Department of Astronomy, School of Physics and Astronomy, Shanghai Jiao Tong University, 800 Dongchuan Road, Shanghai 200240, China; email: jtshen@sjtu.edu.cn, lizy.astro@sjtu.edu.cn\\
$^4$Shanghai Key Laboratory for Particle Physics and Cosmology, 200240, Shanghai China \\
$^5$Department of Astronomy, Tsinghua University, Beijing 100084, China\\
$^6$Key Lab of Optical Astronomy, National Astronomical Observatories, Chinese Academy of Sciences, Beijing 100012, China\\
$^7$Department of Astronomy, The University of Tokyo, 7-3-1 Hongo, Bunkyo-ku, Tokyo 113-0033, Japan\\
$^8$Instituto de Astronom\'ia y Ciencias Planetarias, Universidad de Atacama, Copayapu 485, Copiap\'o, Chile\\}



\begin{abstract}
We revisit the stellar velocity distribution in the Galactic bulge/bar region with APOGEE DR16 and {\it Gaia} DR2, focusing in particular on the possible high-velocity (HV) peaks and their physical origin. We fit the velocity distributions with two different models, namely with Gauss-Hermite polynomial and Gaussian mixture model (GMM). The result of the fit using Gauss-Hermite polynomials reveals a positive correlation between the mean velocity ($\olsi{V}$) and the ``skewness'' ($h_{3}$) of the velocity distribution, possibly caused by the Galactic bar. The $n=2$ GMM fitting reveals a symmetric longitudinal trend of $|\mu_{2}|$ and $\sigma_{2}$ (the mean velocity and the standard deviation of the secondary component), which is inconsistent to the $x_{2}$ orbital family predictions. Cold secondary peaks could be seen at $|l|\sim6\degree$. However, with the additional tangential information from {\it Gaia}, we find that the HV stars in the bulge show similar patterns in the radial-tangential velocity distribution ($V_{\rm  R}-V_{\rm  T}$), regardless of the existence of a distinct cold HV peak. The observed $V_{\rm  R}-V_{\rm  T}$ (or $V_{\rm  GSR}-\mu_{l}$) distributions are consistent with the predictions of a simple MW bar model. The chemical abundances and ages inferred from ASPCAP and CANNON suggest that the HV stars in the bulge/bar are generally as old as, if not older than, the other stars in the bulge/bar region.

\end{abstract}

\section{introduction}
Classical bulges or pseudo-bulges \citep{Kormendy2004}, commonly found in the central regions of disk galaxies, encode essential information about the formation and evolution history of their host galaxies.
Near-infrared observations revealed that the inner regions of the Milky Way (MW) host an asymmetric boxy-bulge \citep{Weiland1994}. Its morphology suggests  that it has a different origin from spherical classical bulges which are likely produced by mergers. \citet{bli_spe_91} postulated that the MW bulge is a tilted bar with the near-end of the bar at positive longitudes. Later studies confirmed that the MW hosts a stellar bar viewed almost end-on \citep{Stanek1994,Nikolaev1997,Unavane1998,Lopez-Corredoira2000,Benjamin2005,Nishiyama2005,Rattenbury2007,Shen2010,Kunder2012,Wegg2013,wegg_15,ness_etal_16, Zasowski2016}. However, many of its structural, kinematical and chemical properties are still under debate \citep{BlaGer2016, Shenli2016, barbuy_etal_18}.

A dynamically cold ($\sigma_{V}\approx$ $30\kms$) high-$V_{\rm GSR}$\footnote{$V_{\rm  GSR}$ is the line-of-sigh velocity with respect to the Galactic Standard of Rest. $\overline{V}_{GSR}$ and $\sigma_{V}$ denote the mean and standard deviation of the high-$V_{\rm GSR}$ peak.} peak ($\overline{V}_{GSR}\approx$ $200\kms$) in the stellar velocity distribution of the Galactic bar/bulge was first reported by \citet{Nidever2012} using the Apache Point Observatory Galactic Evolution Experiment (APOGEE; \citealt{Majewski2017}) commissioning data (DR10). 
The stars contributing to the high-$V_{\rm GSR}$ peak are located at a heliocentric distance of $5-10 \kpc$, indicating that they are not likely associated with the halo or the tidal tails of the Sagittarius dwarf galaxy.
\citet{Nidever2012} concluded that they are possibly bulge/bar stars orbiting in the Galactic bar potential. 
\citet{Molloy2015} suggested that stars in bar-supported 2:1 resonant orbits (e.g. $x_{1}$-orbits) could cause a high-$V_{\rm GSR}$ peak. However, using a self-consistent $N$-body simulation of the Galactic bar, \cite{Lizhaoyu2014} found only a shoulder-like structure instead of a distinct peak at high-$V_{\rm GSR}$. Using simulations which included star formation, \cite{Aumer2015} argued that young stars ($\sim$ 2 Gyrs) on $x_{1}$-orbits captured by the growing bar, may also produce a high-$V_{\rm GSR}$ peak. The $x_{1}$-orbits are the most important type of periodic orbits in almost all numerical models of the galactic bar. However, in models of highly elongated bars the propeller orbits, which are ``distant relatives'' in fact, of the $x_{1}$ orbital family play the dominant role \citep{Kaufmann2005}. \citet{McGough2020} argued that the propeller orbits can be used to explain the high-$V_{\rm GSR}$ peak in three in-plane fields.
In another scenario still, a kpc-scale nuclear disk supported by $x_{2}$-orbits contributes to the formation of the high-$V_{\rm GSR}$ peak \citep{Debattista2015, Fernandez2017}. $x_{2}$-orbits are generally less spatially extended than $x_{1}$-orbits and their major axis is perpendicular to the bar. 

To better distinguish between these scenarios for the high-$V_{\rm GSR}$ (hereafter HV) peaks, in this work we study the bulge kinematics and chemical information based on the APOGEE DR16 and Gaia DR2 surveys. In our previous work \citep{Zhou2017}, we have studied the velocity distributions in the MW bulge/bar using APOGEE DR13. We found 3 fields showing a HV peak at positive longitudes, i.e. fields ($l$, $b$) $=$ $(6\degree,0\degree)$, $(10\degree,-2\degree)$ and $(10\degree,2\degree)$, in the Galactic bulge region. We also found that HV stars show similar chemical abundance ([M/H], [$\alpha$/M] and [C/N])\footnote{These chemical abundances are derived from ASPCAP. [M/H] is the calibrated metallicity.  [$\alpha$/M] is the calibrated $\alpha$-enhancement. [C/N]=[C/Fe]$-$[N/Fe] is the relative carbon to nitrogen ratio.} distributions compared to lower $V_{\rm  GSR}$ stars (the main component in the velocity distribution), indicating that these two components may have a similar age composition. Moreover, in contradiction with previous predictions \citep{Aumer2015}, we found that both young and old stellar populations show HV features.

APOGEE-1 observed only the positive longitude side of the MW bulge/bar. With the goal of better understanding the origin of the HV peaks, in this paper we revisit the stellar kinematics and chemistry of the Galactic bulge/bar using APOGEE DR16 which contains all observed data from APOGEE-1 and APOGEE-2, and covers both positive and negative longitude fields. 

The paper is organized as follows. The sample selection and the stellar parameters estimation are described in \S\ref{sec:data}. The kinematics of the bulge/bar stars is investigated in \S\ref{sec:shapeofvlos}, including the additional information from the radial velocity ($V_{\rm  R}$) and tangential velocity ($V_{\rm  T}$) in  Galactocentric coordinates. In \S\ref{sec:chemical} we compare the chemical abundances and age distributions of different sub-samples. The results are discussed in \S\ref{sec:discussion} and summarized in \S\ref{sec:summary}. 

\section{Data and Method}

\label{sec:data}

\subsection{Sample Selection}

\begin{figure*}[!t]
\centering
\includegraphics[width=1.5\columnwidth]{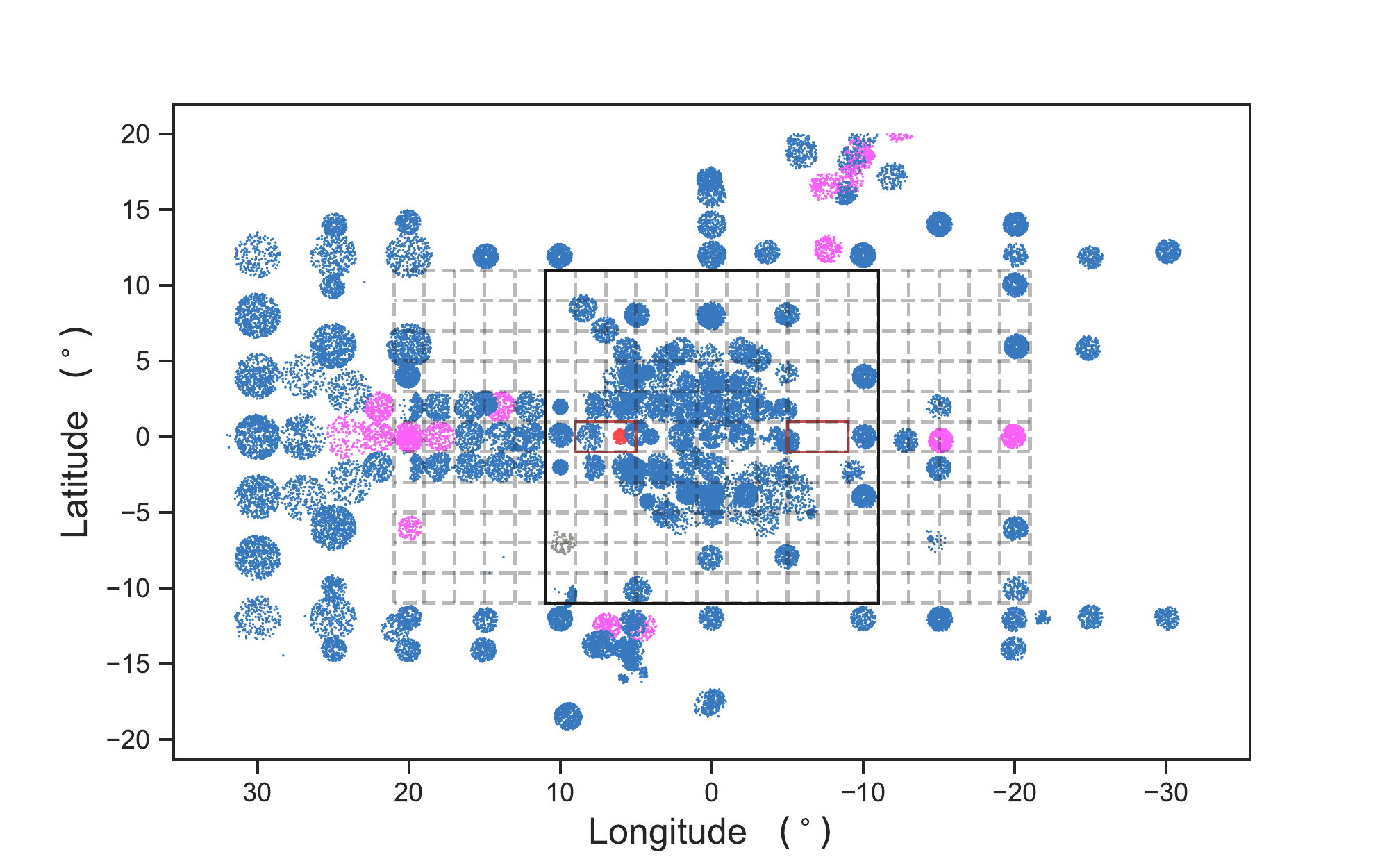}
\centering\caption{Spatial distribution of our main sample (Sample A) from APOGEE DR 16 in the Galactic bulge/bar region (dots). Fields showing a cold HV peak (with a threshold of $\sigma<40\kms$) in the bulge region are marked in red dots. Fields showing a HV peak in the disk dominated region outside the bulge are marked in magenta dots. Two red boxes highlight region with $5\degree < |l| < 9\degree$ and $|b| < 1\degree$, in which HV peaks have been predicted to exist by several theoretical models. The dotted $2\degree \times 2\degree $ grid represents the rebinned fields. The black solid square marks the region of the Galactic bulge. }
\label{fig:1}
\end{figure*}

APOGEE is part of the Sloan Digital Sky Survey IV \citep{Blanton2017}, aiming to map the kinematic and chemical structures of the MW \citep{Majewski2017}. It has built a database of high-resolution (R$\sim$22,500), near-infrared (1.51 to 1.69 $\mu$m H-band) spectra for over $10^{5}$ giant stars. Stars in all the major components of the MW were sampled, including the ``dust-hidden'' parts. APOGEE-1 observed the MW bulge/bar mainly at $l > 0^{\degree}$, whilst APOGEE-2 survey covers the whole bulge/bar region. APOGEE-2 North (APOGEE-2N) and APOGEE-2 South (APOGEE-2S) are two complementary components of APOGEE-2. APOGEE-2N continues observations at Apache Point Observatory using the Sloan 2.5 m telescope \citep{Gunn2006} and original APOGEE spectrograph \citep{Wilson2012}. Meanwhile APOGEE-2S observes the Southern Hemisphere using a cloned APOGEE spectrograph at the 2.5 m Ir\'en\'ee du Pont telescope at the Las Campanas Observatory (LCO) \citep{Bowen1973} . 

{\it Gaia} \citep{Gaia2016, Gaia2018b} is a European space mission to create the most accurate three-dimensional map of the MW. The second Gaia data release ({\it Gaia} DR2) provides high-precision parallaxes and proper motions for 1.3 billion sources down to magnitude G$\sim$21 mag. The proper motions of the cross-matched sample\footnote{{\it Gaia} DR2 cross-matches with APOGEE DR16} have been included in the APOGEE DR16 catalog.
 
From APOGEE DR16 we select our main sample (Sample A) by removing stars with  low signal-to-noise ratio ($\rm SNR < 50$) and high surface gravity (log$(g)> 3.8$) to avoid foreground dwarf contamination. To exclude globular clusters, binaries, and variable stars, we remove stars with large  radial velocity variations during multiple visits, i.e., ${\rm VSCATTER} > 1\kms$ \citep{Nidever2015, Fernandez2016}. In total, Sample A contains $\sim$ 33000 stars within $-20\degree<l<20\degree$ and $-10\degree<b<0\degree$, and we use it in \S \ref{sec:veldisoverall} to investigate the velocity distributions. 

We build a 6D phase-space catalog with the addition of proper motions from {\it Gaia} and stellar distances from StarHorse \citep{starhorse}. Using coordinates transformation we can compute the Galactocentric radius ($R$), azimuth angle ($\phi$), vertical position ($Z$), radial velocity ($V_{\rm  R}$), tangential velocity ($V_{\rm  T}$) and vertical velocity ($V_{\rm Z}$) in the Galactic cylindrical coordinate system. To study the bulge 3D kinematics we select a sub-sample (Sample B) from Sample A which contains both StarHorse distances and {\it Gaia} proper motion measurements. This sub-sample is used to study the proper motion distributions and the correlations between the radial and tangential velocities in \S \ref{sec:vgsr_mul} and \S \ref{sec:vrvt}. After removing stars with proper motion uncertainties greater than 1.5 mas/yr and stars with $d_{\rm err}/d > 30\%$,  $\sim$ 22000 stars are left in Sample B. 

In order to investigate the chemical abundances and the age distributions of the HV stars, we select another sub-sample (Sample C) from Sample A by requiring ASPCAPFLAG=0 for stars with reliable stellar parameters determined by the ASPCAP pipeline. Sample C contains  $\sim$ 22000 stars that are analyzed in \S \ref{sec:chemical} to investigate the chemical and age properties of stars at different velocities.

Following \citet{Zhou2017}, to calculate $V_{\rm  GSR}$, $V_{\rm  R}$ and $V_{\rm  T}$ we adopt the solar position at $(X_{\odot}, Y_{\odot}, Z_{\odot}) = (8.34, 0, 0.027) \kpc$ \citep{reid_etal_14} and the peculiar motion $(U_{\odot}, V_{\odot}, W_{\odot}) = (11.10, 12.24, 7.25) \kms$ \citep{Schonrich2010}, and the circular speed in the Local Standard of Rest of $239\kms$ \citep{McMillan2011}. In the following analysis, we separate stars in each sample into four groups according to their $V_{\rm  GSR}$ and $R$. For example, in the bulge fields with $l > 0^{\degree}$, stars are divided into the disk main component ($R \ge 4 \kpc, V_{\rm  GSR} < 180 \kms$), the disk HV component ($R \ge 4 \kpc, V_{\rm  GSR} \ge 180 \kms$), the bulge main component ($R < 4 \kpc, V_{\rm  GSR} < 180 \kms$) and the bulge HV component ($R < 4 \kpc, V_{\rm  GSR} \ge 180 \kms$). Accordingly, stars with $l < 0^{\degree}$ can also be divided into these four components with $V_{\rm  GSR}$ separated at $-180 \kms$.

We also compare the observational results to a self-consistent $N$-body barred galaxy model from \cite{Shen2010} (hereafter S10). In this model one million particles initially distributed as a pure disk galaxy evolve in a rigid dark matter halo potential. A bar quickly forms and buckles in the vertical direction to form the inner boxy/peanut shaped bulge. The disk substructures become steady after $\sim$2.4 Gyr. As shown in \citet{Shen2010}, this model reproduces the bulge kinematics and morphology remarkably well.

\subsection{Stellar Age Estimation}

To estimate the stellar ages we employ the CANNON, a data driven method to compute stellar parameters and abundances from spectra \citep{Ness2015_cannon,Ness2016_cannon}. Under the assumptions that the continuum-normalized flux is a polynomial of stellar labels and the spectral model is characterized by the coefficient vector of this polynomial, stellar labels are computed by maximizing the likelihood function. \cite{Ness2016_cannon} used 5 stellar labels including effective tempreture $T_{\rm eff}$, surface gravity log$(g)$, metallicity [Fe/H], $\alpha$-enhancement [$\alpha$/Fe] and mass. In our study we tried several sets of stellar labels and selected the set $\{$$T_{\rm eff}$, log$(g)$, [M/H], [$\alpha$/M], [C/Fe], [N/Fe], mass, age$\}$ that provides the best age estimation. We trained a model using age and mass from APOKASC catalog v6.5.4 \citep{Pinsonneault2018} and the corresponding spectra and stellar parameters ($T_{\rm eff}$, log$(g)$, [M/H], [$\alpha$/M], [C/Fe], [N/Fe]) from APOGEE DR16. This sample includes $\sim$6000 stars, with 70\% for training and 30\% for testing. Our test results are shown in the Appendix. The stellar parameters in the test sample are well covered\footnote{ In the bulge region, the log$(g)$ range of our sample (0$-$3.5) is wider than the training sample (1$-$3.5), which may introduce additional uncertainties to the estimated stellar parameters for stars with log$(g)$ less than 1. But it will not statistically change the age distributions of different subgroups. Similar age distributions could be obtained even if we excluded those stars with log$(g) <1$.}, confirming the robustness of our method.
The typical uncertainty of the stellar age is $\sim 0.3$ dex, similar to the results in \citet{Sit2020} using the CANNON on the same training sample (APOKASC). The trained model was then applied to the whole APOGEE DR16 sample.

\section{Kinematic Results}
\label{sec:shapeofvlos}

\begin{figure*}[!t]
\centering
\includegraphics[angle=270,width=1.5\columnwidth]{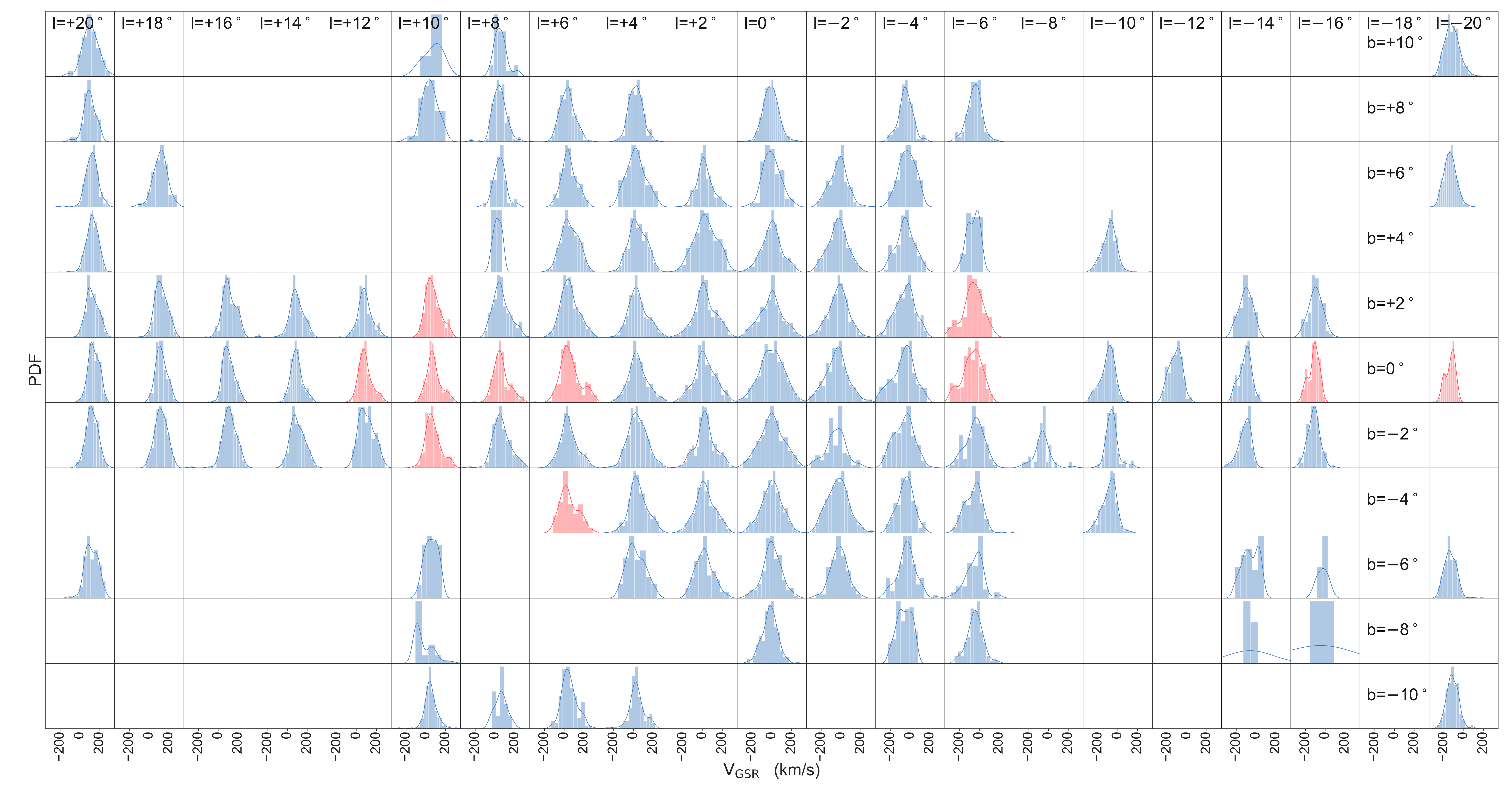}
\centering\caption{$V_{\rm  GSR}$ distributions (including histograms and KDEs) of stars in Sample A within $|l|<20\degree$ and $|b|<10\degree$. Stars are binned as in Fig. \ref{fig:1} (gray dashed grid), with a bin size of $2\degree \times 2\degree$. The velocity distributions showing a HV feature (visual classification) are highlighted in red. Most of them are near the mid-plane.}
\label{fig:2}
\end{figure*}

\subsection{Velocity Distributions}
\label{sec:veldisoverall}

The spatial distribution of Sample A in the Galactic bulge/bar region is shown in Figure \ref{fig:1}. We group the stars in $2\degree \times 2\degree$ bins within $|l|<21\degree$ and $|b|< 11\degree$ (see dashed grid in Fig. \ref{fig:1}). Fig. \ref{fig:2} shows the $V_{\rm  GSR}$ distributions inside the grid, where each sub-panel corresponds to a $2\degree \times 2\degree$ bin. We show the histograms and the kernel density estimations (KDEs). Different choices of bin widths or kernel bandwidth affect the shape of the histograms/KDEs. To balance the sampling noise and at the same time to achieve a better resolution of the density estimation of the velocity distribution, we use the method described in \cite{Freedman1981} to get the optimal bin width and the method in \cite{Silverman1986} to get the optimal kernel bandwidth.

The HV feature is noticeable in several fields (red histograms) near the mid-plane. In most of the bulge/bar fields the velocity distributions show a skewed-Gaussian profile consistent with previous studies, e.g., \cite{Lizhaoyu2014} and \cite{Zhou2017}. There are two types of HV features, namely the distinctive HV peak and the HV shoulder. For the distinctive HV peak, a local minimum is required between the main component and the HV peak in the velocity distribution, as seen in the  $(6\degree,0\degree)$ field. The HV shoulder is usually a skewed Gaussian profile extending towards HV values, e.g., $(8\degree, 0\degree)$\footnote{In \cite{Lizhaoyu2014}, the HV shoulder was not considered as a smaller cold  HV peak enshrouded by the main low velocity peak. The `shoulder' corresponds to the skewed asymmetric shape of the distribution, which may be caused by different mechanisms compared to the high velocity peak. We follow the same definition in this paper}.

\subsubsection{Gaussian Mixture Modeling}
\label{sec:gaussianmix}

\begin{figure*}[!t]
\includegraphics[width=1.8\columnwidth]{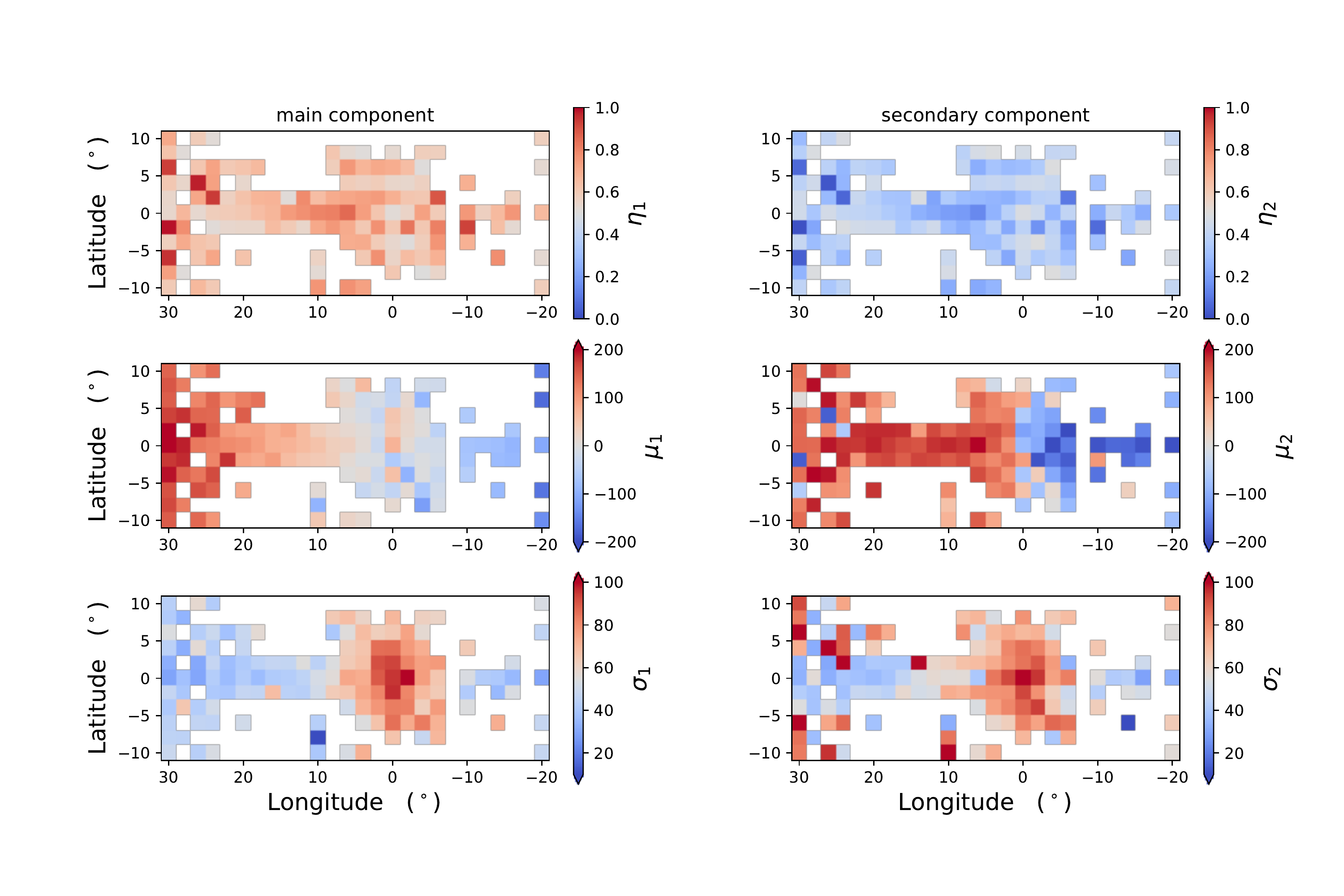}
\centering\caption{Spatial distribution of the weight ($\eta$), mean ($\mu$) and dispersion ($\sigma$) of the GMM fitting ($n = 2$) of Sample A for the main component (left column) and the secondary component (right column). In general the two components show similar longitudinal trend for $\mu$ and $\sigma$, and they also show comparable $\sigma$, except a few bins, e.g., $(6\degree,0\degree)$ and $(-6\degree,2\degree)$. The secondary component corresponds to the HV peak with larger $\mu_{2}$.}

\label{fig:3}
\end{figure*}

\begin{figure}[!t]
\includegraphics[width=1.1\columnwidth]{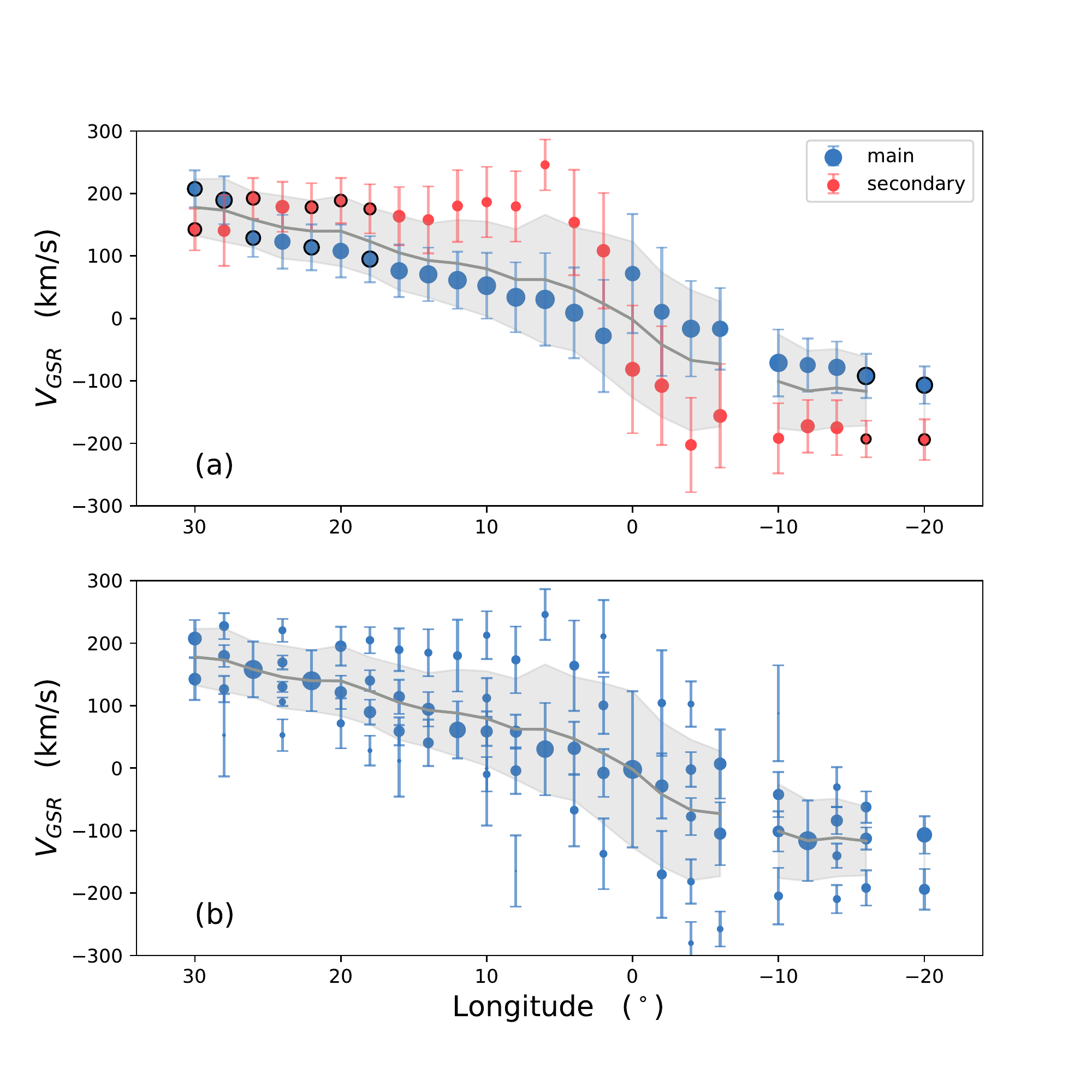}
\centering\caption{The longitudinal profiles of the peak position ($\mu$), velocity dispersion ($\sigma$) and weight ($\eta$) of Gaussian components of GMM fitting results of Sample A. In this figure, the position, error bar, and size of the dots represent $\mu$, $\sigma$, and $\eta$, respectively. The stars are located in the mid-plane, with $-1\degree<b<1\degree$. The shaded region shows the standard deviation of the velocities. (a): GMM with $n=2$ components, where the blue and red points represent the main and the secondary components, respectively. In most bins the secondary peak shows a higher mean velocity. Black circles mark the Gaussians with $\sigma < 40$ km/s. Neither main nor secondary cold peak could be found in $-15\degree<l<15\degree$. (b): GMM with the optimal number of components ($n=1-5$), where $n$ is chosen by the minimum BIC. Usually $3-4$ Gaussians are required in the central region, which indicates that the velocity distributions in the MW central bulge are complicated. This is an additional evidence that there might not be a distinct ``cold HV structure'' in the bulge region.}

\label{fig:4}
\end{figure}

To find the mean velocity and velocity dispersion of the HV feature, we fit the velocity distributions with a Gaussian mixture model (GMM) provided by \texttt{scikit-learn} \citep{scikit-learn} which implements an expectation-maximization (EM) algorithm\footnote{An extra condition $\eta_{n}>0.01$ is adopted. To avoid outliers in GMM fitting we exclude stars with $\left|V_{\rm GSR}\right|>400\kms$. \texttt{scikit-learn} gives random initial coefficient. For every fitting we run the procedure for several times to select the best model with the lowest BIC.}. Since a GMM with 2 components has been commonly adopted by previous studies \citep{Nidever2012, Zasowski2016}, we also initially set the number of Gaussians to $n=$2. A dynamically cold velocity component is defined to have a small velocity dispersion ($\sigma<40\kms$)\footnote{The velocity dispersion $\sigma$ of the cold HV peak reported in \cite{Nidever2012} is around 30$\kms$. In this paper we adopt a slightly higher value (40$\kms$) to ensure the detection of possible cold HV peaks in the bulge region.}. The weight ($\eta$), mean ($\mu$), and standard deviation ($\sigma$) for two Gaussian components are shown in Fig.~\ref{fig:3}. The main components (with $\eta >0.5$) and the secondary component (with $\eta < 0.5$) show a similar longitudinal trend for both $\mu$ and $\sigma$. In most fields, the secondary component has a higher absolute mean velocity than the main component, therefore we dub it the HV component. In the $\sigma_{2}$ distribution we notice a few cold peaks ( $\sigma_{2}\sim40\kms$, light blue) at $|l|\sim6\degree$. 

From Fig.~\ref{fig:1} we can see that the raw observational fields are not uniformly distributed in the bulge area. To reduce the binning effect we also try to group stars by their raw fields and apply the GMM, with the fitting results shown in Fig.~\ref{fig:1}; fields where we find a cold HV component are colored in red inside the bulge region and in magenta outside the bulge region\footnote{Field 010-07-C and some other fields around the Sagittarius Dwarf Spheroidal also show a cold HV peak. We choose not to include those fields in this study since they are probably not related to the Galactic bulge/bar structure \citep[see][]{Zhou2017}.}.

Binning stars by raw fields, in the bulge region $(|l| < 10\degree, |b| < 10\degree)$, with the threshold of $\sigma<40\kms$, only one raw field $(6\degree,0\degree)$ shows a cold peak. Several theoretical works \citep[e.g.,][]{Debattista2015, Aumer2015} predict cold HV features in the disk region delimited by $5\degree<|l|<9\degree$ and $|b|<1\degree$ (see red boxes in Fig. \ref{fig:1}). However, only a few stars were observed in the red box at negative longitudes. We could only use the data in $(\pm5\degree,0)$ and $(\pm10\degree,0)$ to discriminate between different models.

Since all previous models predict the most obvious HV features near the mid-plane ($b= 0^{\degree}$), we show the longitudinal profile of the GMM parameters for the stellar velocity distributions near the mid-plane, in Fig.~\ref{fig:4}.  Only stars within $|b|<1\degree$ are included (the same as the $b=0\degree$ bins in Fig.~\ref{fig:3}). In Fig.~\ref{fig:4}a the number of Gaussians is fixed to $n=2$. The components with $\sigma < 40\kms$ are marked with a black circle, within $|b|<1\degree$ and $|l|<15\degree$ there are no cold HV peaks\footnote{In Fig.~\ref{fig:4}a the $\sigma$ value of the secondary peak in bin $(6\degree,0\degree)$ is slightly greater than that in the raw $(6\degree,0\degree)$ field. Since the $2\degree \times 2\degree$ bin covers a larger area than the raw observational $(6\degree,0\degree)$ field, the inclusion of nearby stars might make the secondary peak hotter.}. 

As shown in Fig.~\ref{fig:3}, the main and the secondary components show comparable $\sigma$ and similar longitudinal trends; in the bulge region ($|l|<10\degree$) the velocity dispersions ($\sigma_{1}$ and $\sigma_{2}$) are larger than that of the disk dominated region ($|l|>10\degree$).

Fixing the number of Gaussians to $n=2$ gives the second simplest GMM but not the best GMM. Therefore we also try to free the number of Gaussians to fit the velocity distribution. The best model giving the lowest Bayesian Information Criterion (BIC) value is adopted. As shown in Fig.~\ref{fig:4}b, at each longitude, the number of points reflects the number of Gaussians $n$ used in the fitting, where $n$ provides the minimum BIC. In the disk region ($|l|>10\degree$) we usually found $n = 1-3$ while in the bulge/bar region ($-10\degree<l<10\degree$) we usually found $n=3-4$. This indicates that the velocity distributions in the MW central bulge are complicated, which could also be considered as an additional evidence that there might not be a distinct ``cold HV structure'' in the bulge region. 

\subsubsection{Gauss-Hermite Polynomials Modeling}

\label{sec:gauss_hermite}
\begin{figure}[!t]
\centering
\includegraphics[width=\columnwidth]{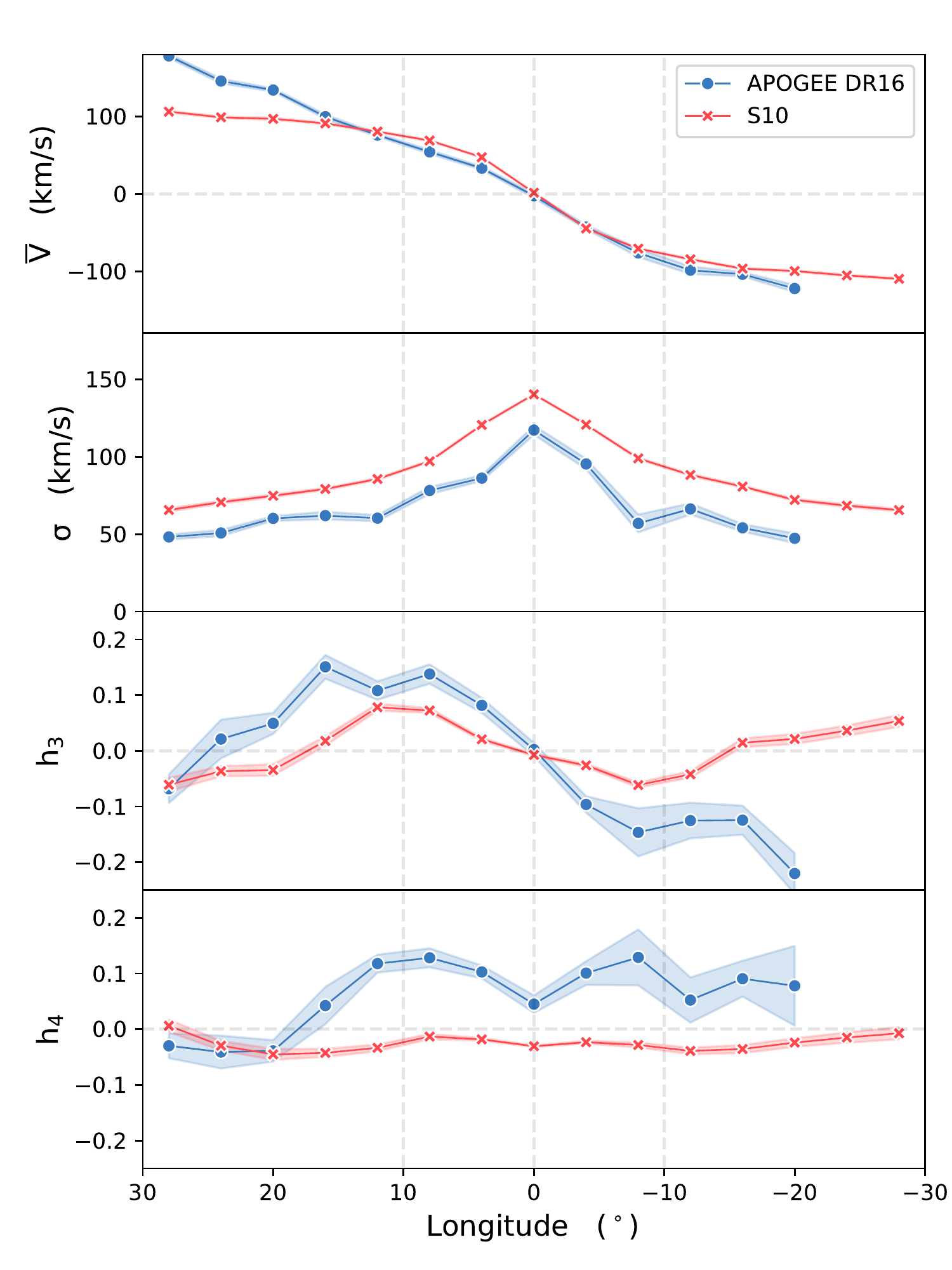}
\centering\caption{Longitudinal profiles of the Gauss-Hermite coefficients: $\olsi{V}$, $\sigma$, $h_{3}$ and $h_{4}$ as shown from top to bottom panels, respectively. Stars in the mid-plane ($-2\degree<b<2\degree$) of Sample A are used here. The shaded region shows the corresponding 1$\sigma$ errors computed using bootstrapping. The observation and the simulation show a similar global trend in $\olsi{V}$, $\sigma$, and $h_{3}$.}
\label{fig:5}
\end{figure}

\begin{figure*}[!t]
\centering
\includegraphics[angle=270,width=1.5\columnwidth]{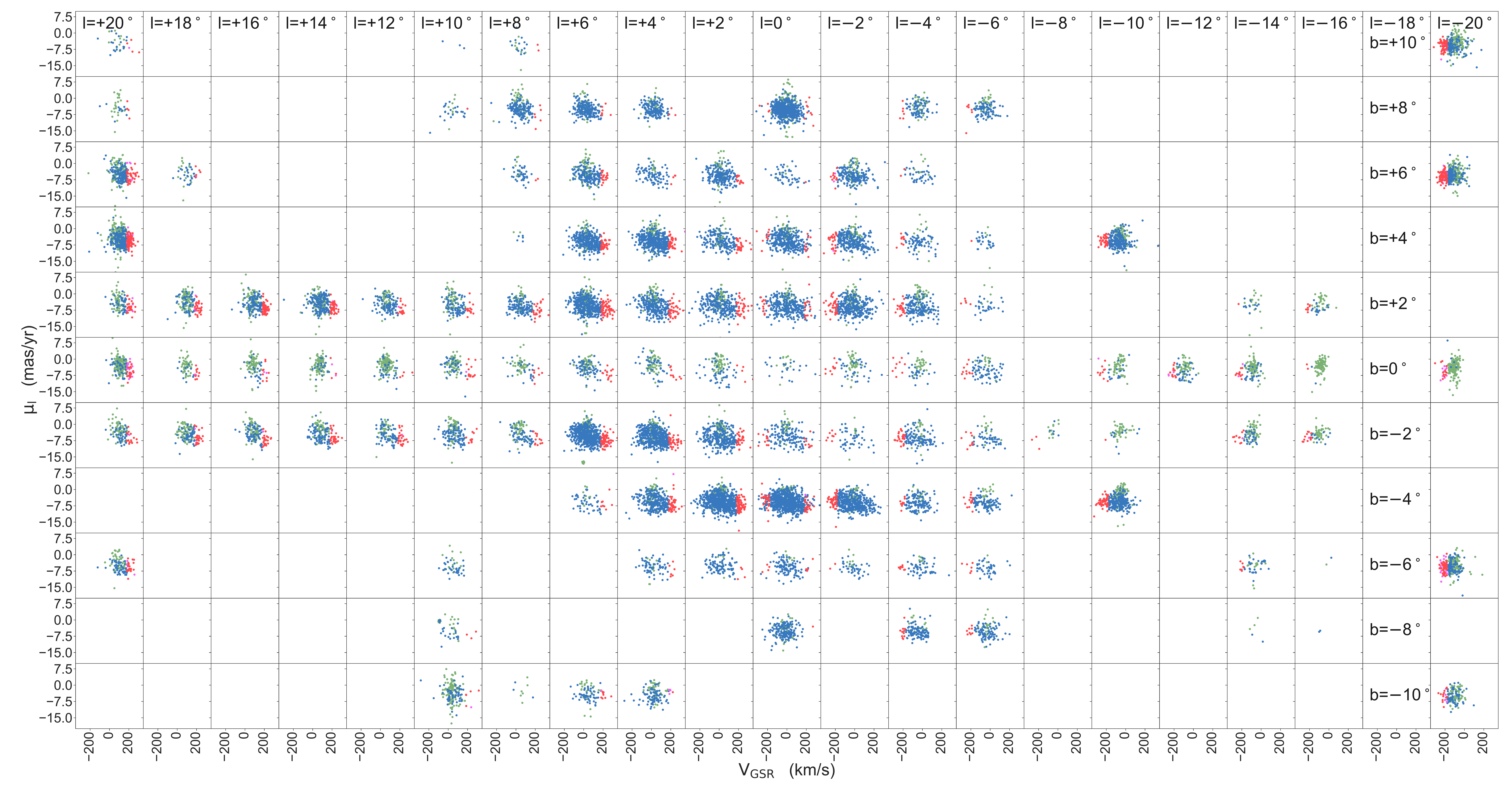}
\centering\caption{
$V_{\rm  GSR}-\mu_{l}$ distributions of stars in Sample B. Green: disk main component; blue: bulge/bar main component; magenta: disk HV component; red: bulge/bar HV component. HV stars generally show different $\mu_{l}$ distribution compared to the main components. 
}
\label{fig:6}
\end{figure*}

In the previous subsection we saw a longitudinal global trend of the velocity and velocity dispersion, suggesting the high-$V_{GSR}$ feature is  possibly related to a global, large scale structure rather than a localized phenomenon.
Gauss-Hermite polynomials are also commonly used to fit quasi-Gaussian profiles \citep{BM1998}, with coefficients $h_{n}$ describe higher order deviations form a Gaussian. We adopt the fitting method in \citet{vanderMarel1993, BM1998}. As it has already been demonstrated in simulations, the relation between the first and the third coefficients ($\olsi{V}$ and $h_{3}$) of a Gauss-Hermite polynomial, which describes the line-of-sight velocity distribution, can be used to characterize the morphology of a galaxy: at large inclination angles these two parameters are positively correlated in the bar region, and negatively correlated in the disk dominated region \citep{Bureau2005, Iannuzzi2015, Lizhaoyu2018}. The $\olsi{V}-h_{3}$ correlation is a useful tool to identify the bar structures in external disk galaxies with IFU observations. The correlation has also been seen in the MW bulge \citep{Zasowski2016, Zhou2017}, although the viewing point is inside the Galaxy at the solar position. 

For a barred galaxy viewed edge-on, the most significant $\olsi{V}-h_{3}$ correlation would show up in the fields close to mid-plane. Considering stars near the mid-plane $(-2\degree<b<2\degree)$, the longitudinal profiles of the four best-fit Gauss-Hermite coefficients $\olsi{V}$, $\sigma$, $h_{3}$ and $h_{4}$ are shown in Fig.~\ref{fig:5} in blue. Sample A is used, groupted by 4$\degree$ $\times$ 4$\degree$ bins, within $32\degree<l<-32\degree$ and $-2\degree<b<2\degree$. Blue circles in the figure mark the central longitude of each bin. For comparison, we computed the same values for the S10 model on the same ($l, b$) grid and the results are shown in red.

The shaded region in the figure shows the statistical errors calculated by the bootstrapping method. Since there are $\sim$ 500 $-$ 2500 stars in most bins, the statistical errors are small. In addition, the stellar position and velocity measurements have small errors, which would not introduce large $h_{3}$ uncertainties. However, some other factors could severely affect the $h_{3}$ measurement, e.g., the incompleteness of the observational data or absence of multiple stellar populations in the S10 model.
Overall, the profiles of the Gauss-Hermite coefficients between the observation and simulation are consistent. In S10, $h_{3}$ reaches its local maximum at $l\sim11\degree$ at positive longitudes and its local minimum at $l\sim-9\degree$ at negative longitudes. These two extrema are slightly asymmetric. This is expected since the turning point in the $h_{3}$ profile might be related to the bar length, and the far-side of the bar ends at smaller $|l|$ than the near-side. The kinematic profiles of APOGEE stars are therefore consistent with theoretical expectations. In the bulge/bar region, $h_{3}$ is positively correlated with $\olsi{V}$. The turning points are also similar to the model; at negative longitudes (far-side of the bar), the turning point of the $h_{3}$ profile appears at slightly lower $|l|$ than the positive longitude (near side of the bar). The low values of $h_{3}$ at $l<-15\degree$ are unexpected but they may be explained by the survey incompleteness at larger distances: in simulations all stars along the line-of-sight are included, while the survey is magnitude limited and may not reach the faint/more distant stars. Generally speaking, considering the variation of the shape of the velocity distribution, the $h_3$ and $\olsi{V}$ profiles and correlation are consistent with predictions of barred models. The HV feature is a natural consequence of a bar.

\subsubsection{Limitations with $V_{\rm  GSR}$ Distributions}

Identifying a dynamically cold feature (in this case the HV component) may shed light on the orbital composition and the substructures of the Galactic bulge. Some earlier works related this cold component to certain resonant orbits \citep{Molloy2015, Aumer2015, McGough2020}, while others argued that this cold component may be a signature of a kpc-scale nuclear disk \citep{Debattista2015}. However, according to our results, the criterion for identifying cold peaks is relatively subjective. The answer to this question heavily depends on the choice of the fitting methods for the velocity distribution, which may vary between different studies. To some extent, enlarging the sample size may not be enough to clarify whether there is indeed a peak in the velocity distribution. Other stellar properties are also needed. Therefore, we consider the additional information of the tangential velocities, chemical abundances and ages in the following sections, which could hopefully provide more constraints on the origin of the HV stars.
\begin{figure*}[!t]
\hspace{-1cm}
\includegraphics[width=2.3\columnwidth]{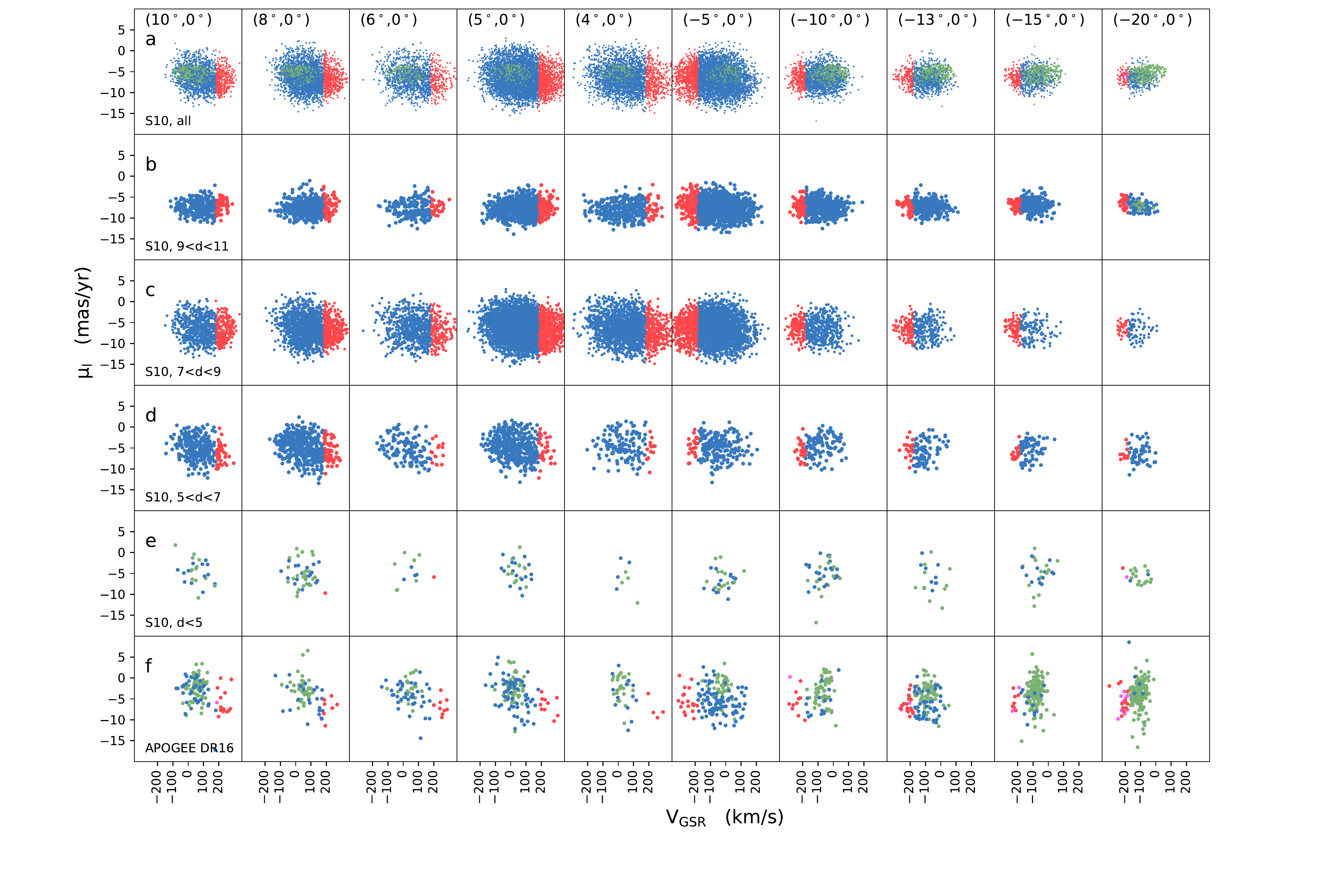}
\centering\caption{
$V_{\rm  GSR}-\mu_{l}$ distributions in some key fields near the mid-plane; (a)-(e) show the simulated $V_{\rm  GSR}-\mu_{l}$ distribution from S10, of particles in different distances. Row (f) shows observational $V_{\rm  GSR}-\mu_{l}$ distribution (Sample B is used here). The longitudes decrease from left to right from $l = 10\degree$ to $-20\degree$. The S10 model distances decrease from top to bottom. In (a) all the particles in the S10 model along the line-of-sight are included. Legends are the same as Fig. \ref{fig:6}. Without selecting stars on particular orbits, the S10 model could reproduce the observed $V_{\rm  GSR}-\mu_{l}$ distributions in most mid-plane fields. The agreement is particularly good if we consider particles in front of the Galactic Center (GC), where most of the APOGEE stars lie, with heliocentric distances between 5 and 7 kpc (see Fig. \ref{fig:7}d).
}
\label{fig:7}
\end{figure*}

\begin{figure*}[!t]
\centering
\includegraphics[angle=270,width=1.5\columnwidth]{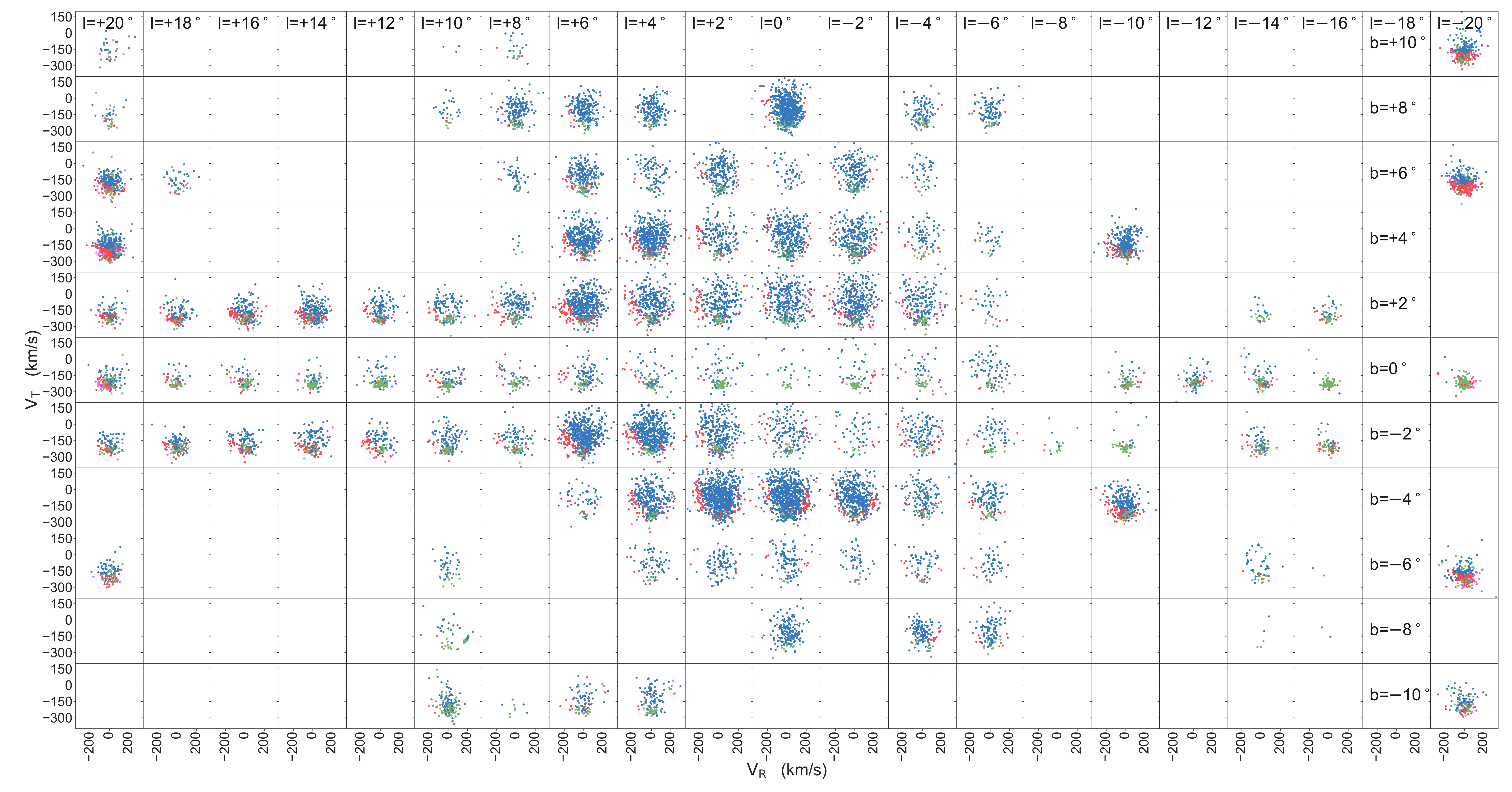}
\centering\caption{$V_{\rm  R}-V_{\rm  T}$ distributions of stars in Sample B in the same region as Fig. \ref{fig:2}, binned by grid.  Legends are the same as Fig. \ref{fig:6}. $V_{\rm  R}-V_{\rm  T}$  distributions of the  HV stars (red) show coherence in different latitudes.}
\label{fig:8}
\end{figure*}

\subsection{$V_{\rm  GSR}-\mu_{l}$ Distributions}
\label{sec:vgsr_mul}

The propeller orbits, ``distant relatives'' of the $x_{1}$-orbits, were recently suggested to contribute to the observed HV peaks \citep{McGough2020}. The propeller orbit family can match the HV peak both in radial velocity and proper motion at $l=4\degree$ (see Fig.~15 in \citealt{McGough2020}). Fig.~\ref{fig:6} shows the distribution of our Sample B in the $V_{\rm  GSR}-\mu_{l}$ space. Each sub-panel of the figure corresponds to a 2$\degree \times 2\degree$ field in the dashed grid of Fig.~\ref{fig:1}. It is apparent that HV stars (red points) at positive longitudes show a larger $|\mu_{l}|$; while at negative longitudes  these HV stars show a smaller/similar $|\mu_{l}|$. These trends are coherent in the whole bulge and are consistent with the trends observed by \citet{McGough2020} for the three fields they studied, at $l = 8\degree, 6\degree$ and $4\degree$.

The simulations, contrary to the observational data, are not affected by observational errors and it is therefore easy to group the particles by their distances. Thus, we use the S10 model to study the distance dependence of the $V_{\rm  GSR}-\mu_{l}$  distributions by grouping stars according to their Galactic coordinates and distances, as shown in Fig. \ref{fig:7}. As mentioned in \S \ref{sec:data}, the stars are divided into disk and bulge components, which are further separated into the HV and the main components. In Fig. \ref{fig:7} these components are labeled with different colors (see caption).

We compare the observations (Fig. \ref{fig:7}f) to the S10 model (Fig. \ref{fig:7}a-e) in the $V_{\rm  GSR}-\mu_{l}$ space, where the S10 model particles are further split into different sub-groups according to their heliocentric distances (e.g. Fig. \ref{fig:7}b contains particles with 9 < $d$ < 11 kpc). We focus on key raw observational fields in the mid-plane of the bulge region, with $|b|<1\degree$ and $|l|>4\degree$ which contain a distinct HV component. Although fields at $l = -13\degree, -15\degree$ and $-20\degree$ are outside the bulge region, we also include them since they are newly observed by APOGEE-2. The particles in the S10 model are selected within the same $(l, b)$ range as the observations. 

Without selecting stars on particular orbits, the S10 model could reproduce the observed $V_{\rm  GSR}-\mu_{l}$ distributions in most mid-plane fields. The agreement is particularly good if we consider particles in front of the Galactic Center (GC), where most of the APOGEE stars lie, with heliocentric distances between 5 and 7 kpc (see Fig. \ref{fig:7}d). The agreement between data and observations is less good in the (4$\degree$, 0$\degree$) field where the APOGEE stars appear to split into two clumps corresponding to the main and HV components as already reported by \citet{McGough2020}. However, the discrepancy between the S10 model and observations in this field could be caused by small number statistics as discussed in the following subsection, where we compare the $V_{\rm  R}-V_{\rm  T}$ distributions between the observations and the model.

Our result seems to indicate a possible connection between the HV stars and the propeller orbit proposed in \citet{McGough2020}, especially in the raw $(4\degree,0\degree)$ field . However, in this field, the velocity distribution of the full APOGEE sample (Sample A) shows no distinct HV peak (see Fig. \ref{fig:2}), which is inconsistent with the prediction of the propeller orbit scenario. More observations with increased number of stars and new fields at negative longitudes are needed to confirm this interpretation for the HV peak.

\begin{figure*}[!t]
\centering
\includegraphics[width=2\columnwidth]{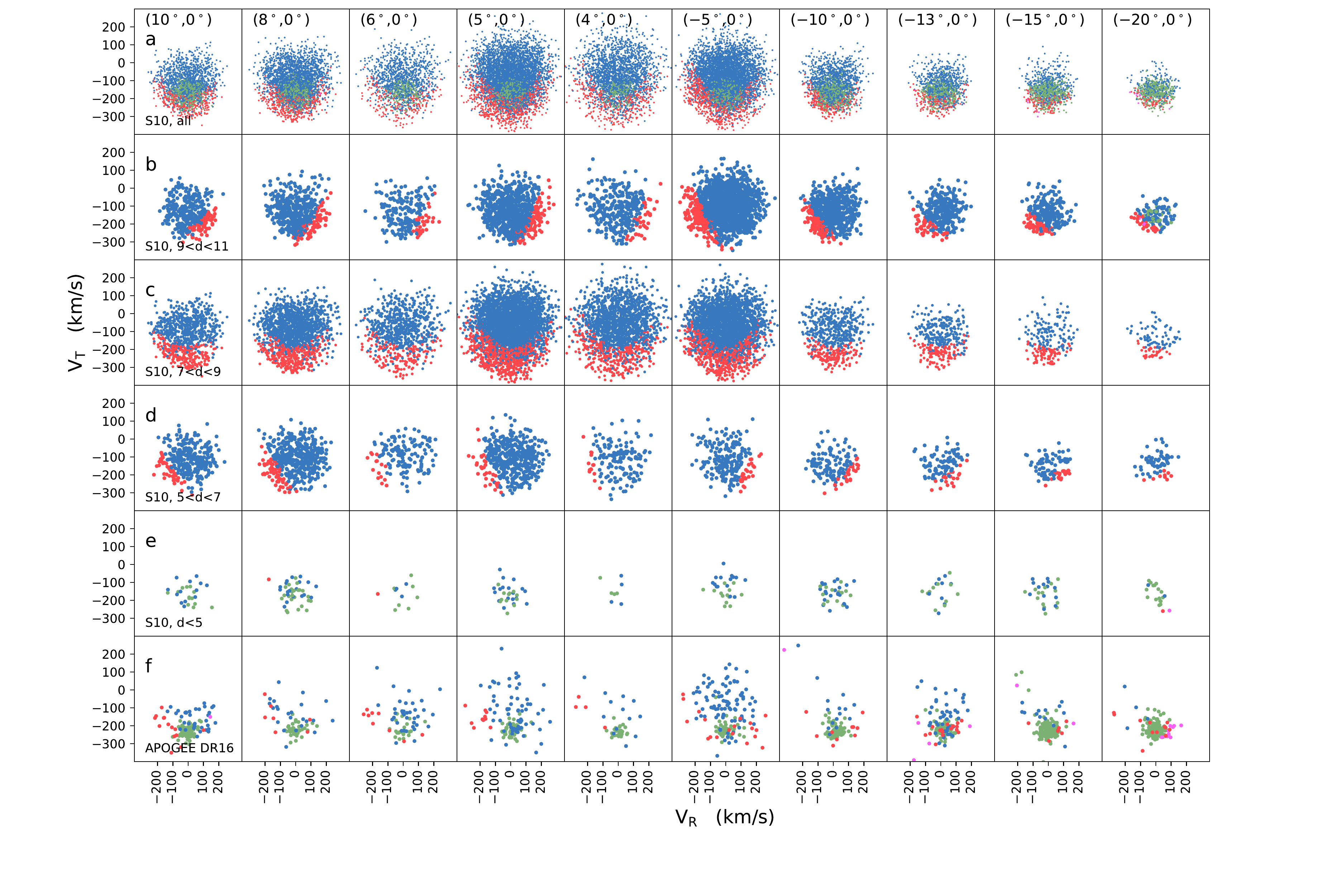}
\centering\caption{$V_{\rm  R}-V_{\rm  T}$ distributions of the same bulge fields in the simulation S10 (row a) and the observed data in Sample B (row f). Rows (b)-(e) are the same as row (a) but only contain stars with different heliocentric distances. The layout and legends are the same with Fig. \ref{fig:7}.} 
\label{fig:9}
\end{figure*}

\begin{figure*}[!t]
\centering
\includegraphics[width=2\columnwidth]{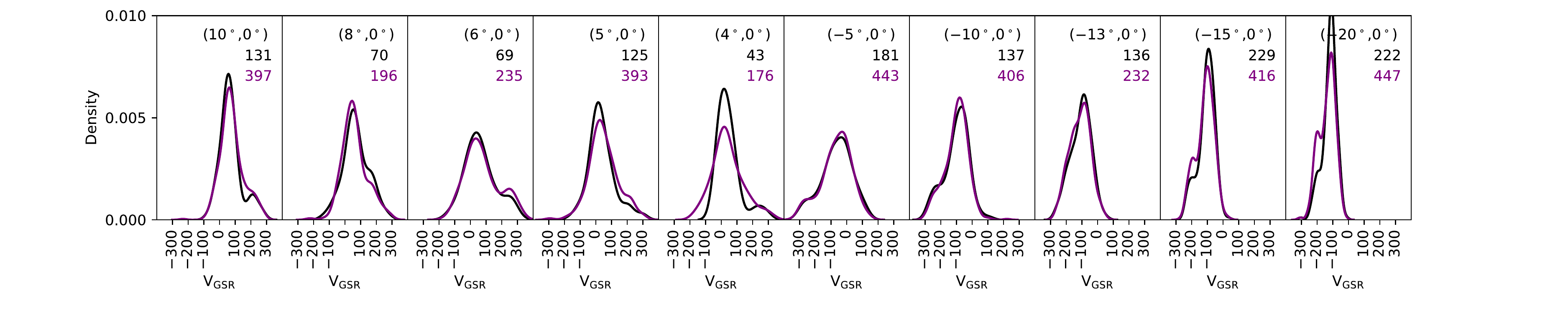}
\centering\caption{Comparison of the $V_{\rm  GSR}$ distributions between Sample A (initial APOGEE DR16 sample, purple lines) and Sample B (APOGEE DR16 sample with {\it Gaia} proper motions and StarHorse distances, black lines). The number of stars in each field is also shown in the corresponding panel. In most of the fields the velocity distributions are similar, indicating that the cross-matching with {\it Gaia} would not change the velocity profiles much. }
\label{fig:10}
\end{figure*}

\subsection{$V_{\rm  R}-V_{\rm  T}$ Distributions}
\label{sec:vrvt}

Using the distance information provided by StarHorse, we can convert the $Gaia$ DR2 proper motions to the velocity space, i.e. we compute the radial velocity $V_{\rm  R}$ and tangential velocity $V_{\rm  T}$ with respect to the Galactic Center. The $V_{\rm  R}-V_{\rm  T}$ distributions are shown in Fig. \ref{fig:8}. In this figure the disk main component (green) shows a smaller velocity dispersion than the bulge/bar main component (blue), as the bulge is generally hotter than the disk. The distributions of HV stars (magenta and red) show longitudinal dependence: in the mid-plane ($b=0\degree$), the HV stars generally have negative $V_{\rm  R}$ for $l > 0\degree$ and positive $V_{\rm  R}$ for $l < 0\degree$. However, in fields above or below the mid-plane, HV stars follow a banana-like distribution in $V_{\rm  R}-V_{\rm  T}$ space in both positive and negative $V_{\rm  R}$.

Both peak-like and shoulder-like HV features could be seen in Fig. \ref{fig:2}. If stars in the peak-like feature have different origins (e.g. due to different kinds of resonant orbits) compared to those in the shoulder-like feature, they may show distinct features in the $V_{\rm  R}-V_{\rm  T}$ space. However, the $V_{\rm  R}-V_{\rm  T}$ distributions of the HV stars in different bulge/bar fields are quite similar, which may imply that HV features share the same origin regardless of the shape of the distributions (i.e. peak-like or shoulder-like).

\begin{figure*}[!t]
\centering

\includegraphics[width=2\columnwidth]{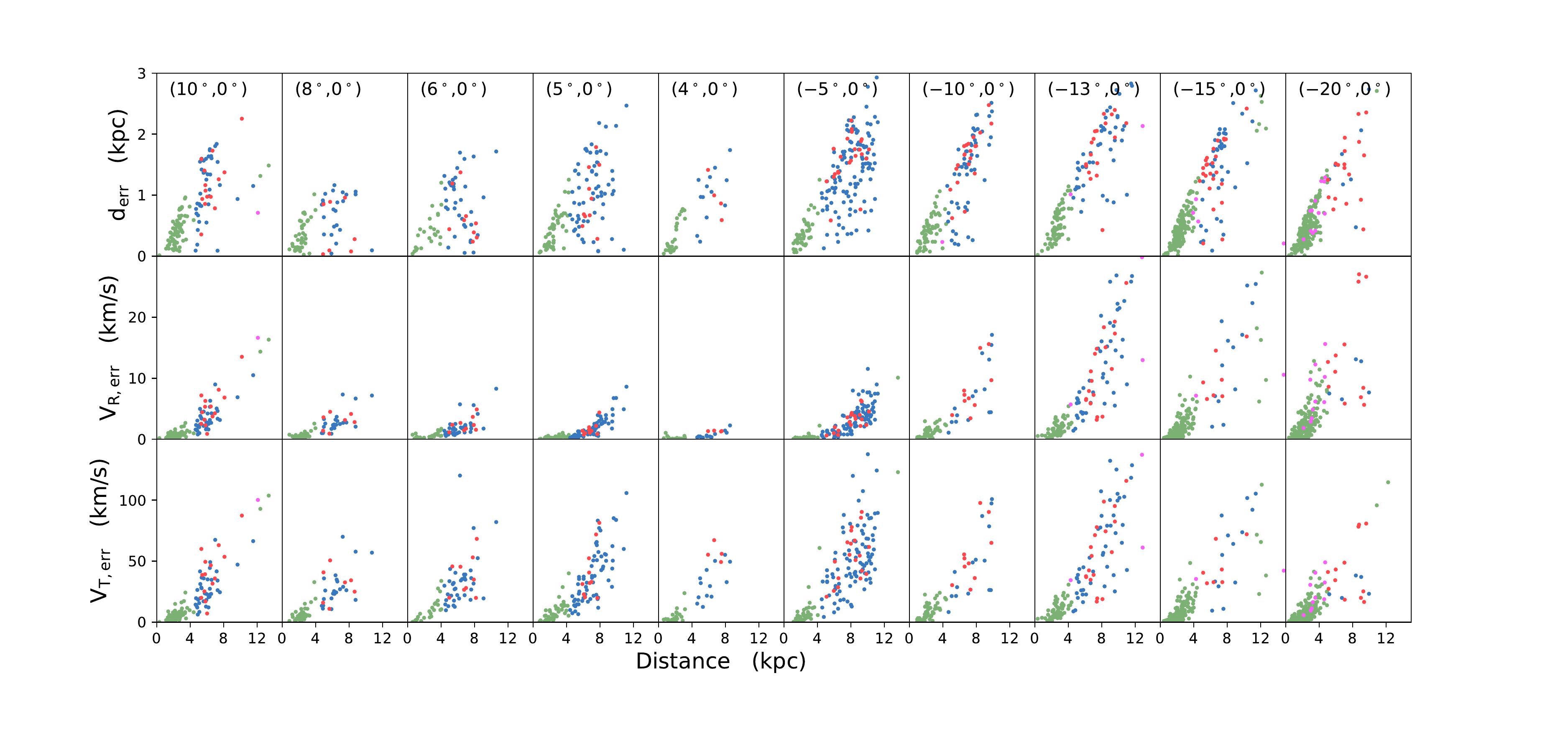}
\centering\caption{Errors in distance, $V_{\rm  R}$, and $V_{\rm  T}$, as a function of distance for Sample B. Legends are the same as Fig. \ref{fig:9}. Errors at positive longitudes are generally smaller than that at negative longitudes.}
\label{fig:11}
\end{figure*}

Fig. \ref{fig:9} is similar to Fig. \ref{fig:7} but now showing the comparison between the observations and the S10 model in the $V_{\rm  R}-V_{\rm  T}$ space. Within the bulge/bar dominated region $|l|<10\degree$, the disk particles have smaller velocity dispersions along the radial and tangential directions than the bulge/bar component. The disk particles constitute a larger fraction of the sample in the fields at $l = -13\degree, -15\degree$ and $-20\degree$ outside the bulge dominated region. In these fields, the disk and the bulge/bar components have similar velocity dispersions. Fig. \ref{fig:9}a, the HV particles showing a banana-shaped distribution can be easily understood since only stars with a large $V_{\rm  R}$ or $V_{\rm  T}$ can contribute to the HV.

As the distance varies, the $V_{\rm  R}$ changes sign because its definition is relative to Galactic Center; as $l$ varies, $V_{\rm  R}$ changes sign because of the symmetry of bar orbits.
For fields at positive longitudes, the location of the HV stars in the $V_{\rm  R}-V_{\rm  T}$ plane (red points) gradually shifts from bottom left to bottom right with increasing distance, with a turning point at around $d=8\kpc$. This could explain the observed different $V_{\rm  R}-V_{\rm  T}$ distribution of the HV stars in Fig. \ref{fig:8}. In mid-plane fields, most HV stars appear only on one side (at positive or negative $V_{\rm  R}$). However, in off-plane fields the distribution of HV stars becomes roughly symmetric and banana-like and they can appear at both positive and negative $V_{\rm  R}$. This is probably due to stronger dust extinction in the mid-plane, which obscures distant stars.

Fig. \ref{fig:9}f shows the APOGEE $V_{\rm  R}-V_{\rm  T}$ distributions in the same fields as the other rows.  Similar to Fig. \ref{fig:9}d, HV stars at positive longitudes generally show negative $V_{\rm  R}$. This is consistent with our arguments that at positive longitudes most of the HV stars are within 8 kpc from the Sun, while at negative longitudes the distance distributions are broader and no clear trend can be observed.

In Fig. \ref{fig:6} and Fig. \ref{fig:9} Sample B is used, as a sub-sample of Sample A.  Sample B contains only stars with $Gaia$ DR2 proper motions and StarHorse distances, and acceptable errors on proper motion and distance. In Fig. \ref{fig:10} we compare the $V_{\rm  GSR}$ distributions of Sample A (purple) and Sample B (black). In most fields, they are very similar, confirming that Sample B is not strongly affected by selection effects. Stars in Sample A could reach a StarHorse distance of 10 kpc. In most fields, the distance distribution of Sample B does not change much and can still reach 10 kpc, indicating that Sample B extends far enough to reach the bulge/bar also at negative longitudes.

The $V_{\rm  R}-V_{\rm  T}$ distributions of the S10 model at $l = -15\degree$ and $-20\degree$ shown in Fig. \ref{fig:9}a only contains a small fraction of bulge/bar stars (blue and red), with the majority as disk stars (green and magenta). The corresponding APOGEE fields in Fig. \ref{fig:9}f also show that the disk is the dominant population at these longitudes. Therefore, the HV feature in these two fields might not be related to the bulge/bar. Sample A in these two fields shows a HV peak.  After cross-matching with {\it Gaia} and removing stars with large distance uncertainty, the HV components in the velocity distributions become less noticeable as shown in the right two panels in Fig.~\ref{fig:10}. This means that after the cross-match more HV stars may have been removed from the sample compared to the main component. Therefore it is difficult to draw any further insights on the origin of the HV stars from the $V_{\rm  R}-V_{\rm  T}$ distributions.

\subsection{Uncertainties}

When studying the $V_{\rm  R}-V_{\rm  T}$ distributions we exclude stars with large proper motion errors ($>$1.5 mas/yr) and large relative distance errors ($d_{\rm err}/d>0.3$). The proper motion uncertainties and the cross-terms are provided by $Gaia$ DR2, the distance errors by StarHorse and the radial velocity errors by APOGEE-2. We compute the uncertainties for the Galactocentric velocities along the radial and tangential directions, propagating the covariance matrix from one coordinate system to another. 

For the same sub-sample in Fig. \ref{fig:10}, we show the heliocentric distance errors from StarHorse and the newly derived $V_{\rm  R}$ and $V_{\rm  T}$ errors in Fig. \ref{fig:11}, as a function of StarHorse distance.
Within $|l|<5\degree$, the $V_{\rm  R}$ errors are small since they are mainly contributed by the $V_{\rm los}$ errors which are very small. With increasing $|l|$ , the $V_{\rm  R}$ uncertainties become larger, reaching up to 30 $\kms$. The $V_{\rm  T}$ errors show the opposite trend: for small $|l|$, $V_{\rm T,err}$ are large. 
 
From the analysis of the velocity errors we can conclude that $V_{\rm R, err}$ is acceptable in the bulge region. Note that velocity errors at positive longitudes are smaller than at negative longitudes, which might also explain why clear trends of HV stars in the $V_{\rm  R}-V_{\rm  T}$ distributions could only be seen in the near side of the bar. 

\section{chemical abundances and age distributions of HV stars}
\label{sec:chemical}

\begin{figure*}[!t]
\centering
\includegraphics[width=2\columnwidth]{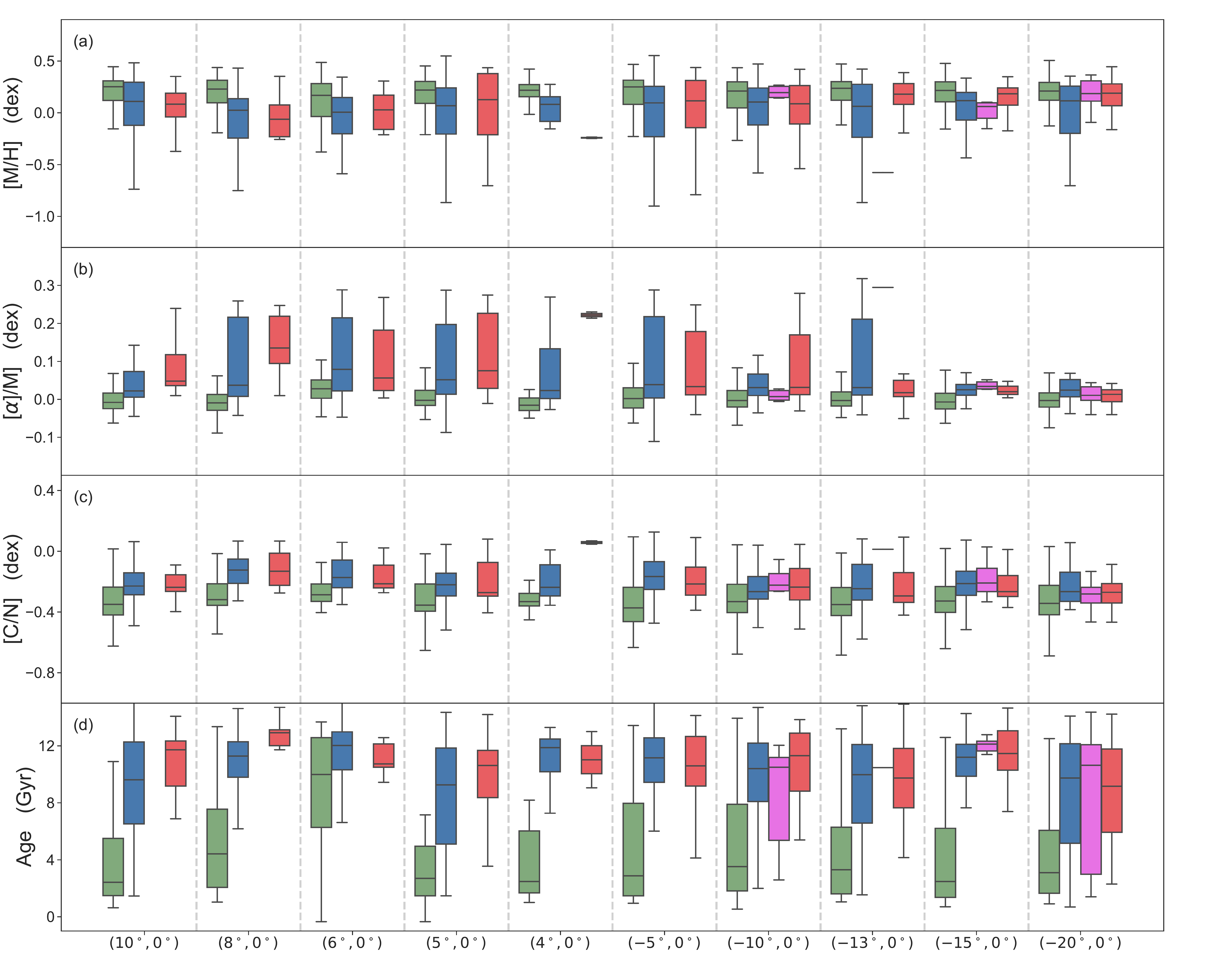}
\centering\caption{Box plots of the chemical abundances and ages of the four components for Sample C in the same fields as those in Fig. \ref{fig:9}. Distributions of [M/H], [$\alpha$/M], and [C/N] are shown in (a), (b) and (c) respectively. Green: disk main; blue: bulge main; magenta: disk HV; red: bulge HV. The box shows the quartiles ($25\% - 75\%$) of the dataset while the whiskers extend to show the rest of the distribution. Similar distributions (both chemical abundance and age) could be found between the bulge main component (blue) and the bulge HV component (red).}
\label{fig:12}
\end{figure*}

In this section we study the chemical abundance distributions of bulge/bar stars at different velocities in fields along the mid-plane ($b=0\degree$). As mentioned in $\S$ \ref{sec:data} we selected Sample C from the main Sample A with the ASPCAP stellar parameters available (ASPCAPFLAG=0). Stars are then grouped by their $V_{\rm  GSR}$ and Galactocentric radius $R$ and further divided into four components (i.e., disk main, bulge main, disk HV and bulge HV components) as described in $\S$ \ref{sec:data}.

Similar to \cite{Zhou2017}, we also use the chemical abundances ([M/H], [$\alpha$/M] and [C/N]) as age proxies. The metallicity and $\alpha$-abundance trace the age from the chemical evolution aspect, while [C/N] traces the age from the stellar evolution aspect \citep{Martig2015}. Massive stars (which are younger in general) would have a lower [C/N] after their first dredge-up phase during the stellar evolution process. This relation is valid for our sample as most stars have experienced their first dredge-up phase since their log$\,(g)<3.5$. A recent study by \cite{Shetrone2019} argued that for metal-poor ([Fe/H]<$-$0.5), evolved (log$(g)<2$) red giants, [C/N] may not be a reliable age indicator, but our conclusion here should not be affected since such stars only contribute a small fraction to our sample ($\sim$4\%).

To study the chemical properties of stars with different velocities, we select stars from Sample C. For the same fields in Fig. \ref{fig:9} we plot the distributions of the three chemical parameters of the four components (see Fig. \ref{fig:12}). Our sample covers a large distance range (up to $10$ kpc) from the Sun. However, bulge stars (with Galactocentric radius $R<4$) only contribute a small fraction in the sub-sample. There are more metal poor or $\alpha$-enhanced stars in the bulge/bar fields ($|l|<10\degree$) compared to the disk fields.

Velocity distributions in the raw field $(6\degree,0\degree)$ shows the coldest HV peak to others in the bulge/bar region. The chemical distribution of stars in this field is no different to that of the other bulge fields within $|l|<10\degree$ (see Fig. \ref{fig:12}). HV stars at $(4\degree,0\degree)$ show quite different chemical abundance, but the number of stars in this field is small. A larger sample with more accurate kinematical and chemical information is needed to better understand the origin of HV stars in this field. Generally speaking, in the bulge region $(|l|<10\degree)$, the bulge/bar components (blue and red) have relatively low metallicity, high [$\alpha$/M], and high [C/N] compared to the disk components (green and magenta). We find no significant difference in the chemical abundance distributions of the bulge/bar main component and the HV component, indicating that they belong to a similar stellar population, consistent with \citet{Zhou2017}.

As mentioned in \S 2.2, we have estimated the stellar ages from the CANNON pipeline. Fig. \ref{fig:12}d shows the direct age comparison between the four components. In most of these fields, stars in the disk components (green and magenta) are relatively younger than those in the bulge/bar components (blue and red). On the other hand, the age distributions of the two bulge/bar components (blue and red) are similar. This result is consistent with the conclusion of our previous work. In the bulge/bar region, there is no clear age difference between stars in the HV peak and the main component \citep{Zhou2017}.

\section{Discussion}
\label{sec:discussion}

\subsection{Comparison with Previous Results}

The identification of a secondary peak in the velocity distribution is not straightforward. Different methods have been utilized in previous studies. For example,  \cite{Nidever2012} used a double-Gaussian model to fit the velocity distributions to extract a secondary peak at high velocity with small velocity dispersion in several bulge fields. In our previous work, \cite{Zhou2017} measured the strength of the HV peak/shoulder by calculating the deviation of the velocity distribution from a Gaussian profile. Here, besides the Gauss-Hermite approach, we also fitted the data with a GMM with \texttt{scikit-learn} in \S\ref{sec:gaussianmix}.  In \cite{Lizhaoyu2014} the HV shoulder is not considered as a  weaker Gaussian enshrouded by the main one but was  described as a skewed asymmetric shape of the distribution. There might be different mechanisms between the HV peak and the HV shoulder, which may related to the formation and evolution history of the Galactic bulge. Comparing distributions in other dimensions, i.e., $V_{\rm R}-V_{\rm T}$ and chemical abundance distributions, we find no obvious difference between the fields showing a shoulder-like and a peak-like HV feature. Thus HV peaks and shoulders might not be essentially different. Instead, it is more important to confirm whether the HV feature is cold. If we chose $\sigma<40\kms$ as a threshold for a cold peak inside the bulge/bar, then the secondary peaks in most regions would not be cold (see Fig. \ref{fig:1} and Fig. \ref{fig:3}).

\cite{Nidever2012} reported several fields showing a cold HV peak at positive longitudes ($4\degree<l<14\degree$), with a spatial distribution slightly asymmetric - with colder HV peaks at negative latitudes. At negative longitudes HV peaks have also been reported at $(-6\degree, 0\degree)$ \citep{Babusiaux2014} and $(-8.5\degree, 0\degree)$ \citep{Trapp2018}. On the other hand, the HV features could be widely spread in the bulge/bar region. Although the two fields mentioned above are not covered by APOGEE DR16, the general longitudinal trend of the secondary peak is consistent with these previous studies.

The longitudinal trend of the velocity distributions is also consistent with previous works and model predictions of bar kinematics. As shown in Fig. \ref{fig:4}, at $5\degree<|l|<20\degree$ the secondary component in the GMM, i.e., the HV feature, shows similar mean values ($\mu \sim 200\kms$), which is consistent with \cite{Zasowski2016}. Fitting the profiles with Gauss-Hermite polynomial the correlation between its coefficients $\olsi{V}$ and $h_{3}$ is also consistent with the theoretical expectations of a bar structure \citep{Bureau2005, Iannuzzi2015, Lizhaoyu2018}.

\cite{Zhou2017} found no chemical abundance difference between stars in the HV peak and the main component, which implied that their age composition is similar. By comparing the updated chemical abundances [M/H], [$\alpha$/M] and [C/N] for stars in more fields we still find no distinction between these two components. Further more, the age distribution in Fig. \ref{fig:12} also supports the conclusion that these two component have a similar age composition.

The present work provides significant improvements over our previous work.  \citet{Zhou2017} used APOGEE DR13, which mainly covers the positive longitudes side of the bulge/bar. This work uses APOGEE DR16, which contains more fields at negative longitudes. The full bulge coverage is very important in discriminating between different model predictions. Moreover, we estimate the stellar ages using the CANNON pipeline which enables us to study the kinematic properties of stars in different age ranges.  We also utilized the Gaia data and the StarHorse catalog to get robust distances and 6D phase-space information for stars in the bulge region. Therefore, we can compare the data with model predictions in new phase-space, e.g., the $V_{R}-V_{T}$ space.

\subsection{Metallicity and Age bias}

As shown in Fig. \ref{fig:12}, the bulge main component (blue) and the bulge HV component (red) have similar chemical abundances and age distributions. In most fields, the foreground disk components (green and magenta) are slightly more metal-rich than the bulge/bar components (blue and red), implying a global positive radial metallicity gradient from the inner region to the outer disk. Such positive metallicity gradient, different from the weak negative radial metallicity gradient revealed by GIBS and ARGOS \citep{Ness2013, Zoccali2017}, was also reported in \cite{Fragkoudi2017} using APOGEE DR13. The positive radial metallicity gradient might be due to the potential metallicity bias discussed in \cite{GarciaPerez2018}. In the spectroscopic analysis due to the cool edge of the model grid (ASPCAP, $T_{\rm eff} \sim 3500$K), some metal-rich old stars were missing. Metal-rich stars in the MW disk are usually young so this selection effect makes bulge stars more metal-poor on average. For the same reason stars in the red giant branch are better represented at younger ages statistically \citep[see Section 3 in][]{GarciaPerez2018}, which is also the case of our Sample C shown in Fig. \ref{fig:12}. However, the missing old metal-rich stars should still be included in Fig. \ref{fig:2} with Sample B. Therefore, no such age bias should exist in the velocity distributions.

\subsection{The Origin of the HV Component}

Several models were proposed to explain the HV peak. A simple conjecture of its origin is stars in the resonant orbits. The HV peaks might have a gaseous counterpart, as HV peaks could also be seen in the $(l,V_{\rm los})$ distributions, which are usually interpreted as gas following $x_{1}$-orbits \citep {Sormani2015,Lizhi2016}.
As the most populous orbits in barred potentials \citep{SparkeSellwood1987}, \cite{Molloy2015} showed that the $x_{1}$ orbital family is able to simultaneously generate peaks close to the observed value. This scenario is also supported by \cite{Aumer2015}. With a barred model containing star formation, they further pointed out that at a given time, $\sim40-50$ percent of HV stars are in $x_{1}$-like orbits and they are preferentially young, since new born stars are easily trapped by $x_{1}$-orbits. Recently, \citet{McGough2020} suggested that the propeller orbit, as a ``distant relative'' of the $x_{1}$ family, could also be responsible for the observed HV peaks in the MW bar, and the fraction of propeller orbits could provide constrains to MW modeling. In models of very elongated bars, propeller orbits are common and play a very dominant role \citep{Kaufmann2005}.\footnote{Connections and differences between the $x_{1}$ and the propeller families could be found in Fig.~6 of \citet{Kaufmann2005}.} The $x_{2}$-orbits belong to another important resonant orbital family in the bar region. \cite{Debattista2015} suggested that a kpc-scale nuclear disc (or ring), composed of stars on $x_{2}$-orbits, is responsible for the HV peak. 

\begin{table}[h]
  \begin{center}
    \caption{GMM fitting (n=2) parameters}
    \label{tab:table1}

    \begin{tabular}{c|ccc|ccc} 
      region  & $\eta_{2,1}$ & $\mu_{2,1}$ & $\sigma_{2,1}$ & $\eta_{2,2}$ & $\mu_{2,2}$ & $\sigma_{2,2}$\\
  
      \hline
      $(-5\degree,0\degree)$ & 0.68 & $-$18.83 & 69.64 & 0.32 & $-$175.34 & 83.14 \\
      $(5\degree,0\degree)$ & 0.75 & 13.62 & 71.62 & 0.25 & 167.85 & 84.69 \\
      $(-10\degree,0\degree)$ & 0.75 & $-$71.22 & 53.59 & 0.25 & $-$191.91 & 56.04 \\
      $(10\degree,0\degree)$ & 0.80 & 52.37 & 52.54 & 0.20 & 186.20 & 56.35 \\
    \end{tabular}
  \end{center}

 { Note: The first footnote shows the total number of Gaussians. The second footnote `1' and `2' show the main component and the secondary component, respectively. At negative longitudes the secondary component show similar $\sigma$ and slightly larger $|l|$.}

\end{table}

\begin{table}[h]
  \begin{center}
    \caption{GMM fitting (n=3) parameters of the weakest component}
    \label{tab:table2}
    \begin{tabular}{c|ccc} 
      region & $\eta_{3,3}$ & $\mu_{3,3}$ & $\sigma_{3,3}$ \\
  
      \hline
      $(-5\degree,0\degree)$ & 0.19 & $-$227.77 & 55.91 \\
      $(5\degree,0\degree)$ & 0.20 & 200.88 & 63.20\\
      $(-10\degree,0\degree)$ & 0.20 & $-$209.87 & 44.25 \\
      $(10\degree,0\degree)$ & 0.19 & 196.60 & 46.06 \\

      \end{tabular}
      \end{center}
     {Note: This component (the third component) has the highest $V_{\rm GSR}$ among the three Gaussians. Similar to the n=2 GMM fitting, at negative longitudes the weakest component show similar $\sigma$ and slightly larger $\left|\mu\right|$.}
\end{table}

We could use the observational data to discriminate between these models, since the $x_{1}$ and the $x_{2}$ scenarios predict distinct secondary peaks at different longitudes with different $\mu$ and $\sigma$. \cite{Aumer2015} predicted  HV peaks at $l=3\degree - 10\degree, |b|<1\degree$. In their Fig.~4, the velocity profiles at the same $|l|$ are almost symmetric, i.e., have the same $|\mu|$ and $\sigma$. At positive longitudes, \cite{Debattista2018} predicted HV peaks at $l=8\degree - 12\degree$, also in the mid-plane. As seen in their Fig.~7, HV peaks at negative longitudes show smaller $|\mu|$ and smaller $\sigma$. 
Although some key fields are not covered by APOGEE-2, we still have the $(\pm5\degree, 0\degree)$ and $(\pm10\degree, 0\degree)$ fields to discriminate between $x_{1}$ and $x_{2}$-orbits scenarios. Tab.~\ref{tab:table1} and Tab.~\ref{tab:table2} show the GMM fitting results for $n=2$ and $n=3$ respectively. The velocity dispersions $\sigma$ of the HV component ($\sigma_{2,2} $ for $n=2$ and $\sigma_{3,3}$ for $n=3$) are similar at the same $|l|$. The peak velocities $|\mu_{2,2}|$ and $|\mu_{3,3}|$ at negative longitudes are slightly larger than those at positive longitudes. This symmetry could also be seen in Figures \ref{fig:3} and \ref{fig:4}. This observational result is inconsistent with the \cite{Debattista2015} model predictions that the HV peaks at negative longitudes should have smaller peak velocity and velocity dispersion.

The propeller orbit, a ``distant relative'' of the $x_1$ orbit, has also been suggested to contribute to the observed HV peaks in the MW bar \citep{McGough2020}. They argued that this kind of orbits could explain the observed proper motions of stars in fields at $ l = 4\degree, 6\degree, 8\degree$ in the mid-plane $(b=0\degree)$. To make a direct comparison with their results, we plot the same $V_{\rm  GSR}-\mu_{l}$ distributions of stars in several key fields in the mid-plane. The S10 model is also included for comparison (see Fig. \ref{fig:7}). In this figure, S10 could match the observational data in most fields, except for $(4\degree, 0\degree)$, where the distribution shows a large slope and is clearly divided into two clumps. 
In the same raw field we also found that the distances of HV stars are concentrated at $d=6.5\kpc$. Interestingly, according to Fig. 15 in \cite{McGough2020}, $l=4\degree$ is close to the line-of-sight tangential point of the propeller orbit at a distance of $\sim7\kpc$ from the Sun, indicating a possible connection between the HV stars at (4$\degree$, 0$\degree$) with the propeller orbit.

As shown in Fig. \ref{fig:8}, HV stars in different bins show a latitudinal similarity and a clear longitudinal trend in the $V_{\rm  R}-V_{\rm  T}$ distributions, regardless of the shapes of velocity profiles. The good agreement between the self-consistent barred model S10 and the APOGEE observations further supports the tight relation between the bar and the HV stars. The $(4\degree,0\degree)$ field seems to be an exception, but in this field the number of HV stars is still too small to be conclusive. 

Generally speaking, the HV feature has a tight relation with the bar but may not be necessarily related to some special resonant orbits, since in S10 all the particles in the simulation within a certain radius range ($5<d<7\kpc$) are included without selecting any particular type of orbits.

In addition to kinematics, chemical abundances and age information can help to further constrain the origin of the HV peak. \citet{Aumer2015} suggested that young stars with age $<2\Gyr$ contribute to a HV peak in the velocity distribution. In this scenario the fraction of young stars is higher in the HV peak than the main component, since both young and old stars could contribute to the main component, but only young stars contribute to the HV peak. The nuclear disk scenario in \citet{Debattista2018} also implies that stars on the $x_2$ orbit are generally younger than the bulge/bar main component, since the nuclear disk forms later inside the bar structure. In this study, with more fields, better derived chemical abundances and estimated age from the CANNON, we still find no evidence for younger stars in the HV component than the main component, as shown in Fig.~\ref{fig:12}, consistent with \citet{Zhou2017}.

The raw field $(6\degree, 0\degree)$ shows the coldest secondary peak in $V_{\rm GSR}$ distribution. However, the $V_{\rm  R}-V_{\rm  T}$ distribution, chemical abundance distribution and the age distribution of stars in this field show a similar trend to stars in the other fields. Thus it might not have a unique origin. We also checked the globular clusters distribution in the inner galaxy \citep{Perez2020} and found no corresponding globular clusters near the $(6\degree,0\degree)$ field. 

Using the BIC to select the best number of Gaussians $n$ in the GMM, we find that $n = 3-5$ Gaussians are usually needed in the bulge/bar region, indicating that velocity distributions in the bulge/bar region are relatively complicated. More observations are needed in the future to verify the existence of the cold HV peaks in the bulge. The number of stars in each field should be increased  to achieve better statistics. In addition, more spatial coverage at negative longitude is necessary. For example, $(4\degree, 0\degree)$ and $(-6\degree,0\degree)$ are the two key fields to distinguish between different scenarios. However, the sample size in the $(4\degree, 0\degree)$ field is relatively small, and the $(-6\degree, 0\degree)$ field even lacks observational coverage. With more targets observed and larger spatial coverage in the bulge region, predictions from different scenarios could be better verified.

\section{Summary and Conclusion}
\label{sec:summary}

We revisit the stellar velocity distributions in the bulge/bar region with APOGEE DR16 and {\it Gaia} DR2 , focusing on the possible HV peak and its physical origin. Our sample covers the Galactic bulge in both the positive and negative longitudes, allowing for a comprehensive analysis of the Galactic bulge. We adopt StarHorse distances and {\it Gaia} proper motion to derive the radial velocity ($V_{\rm  R}$) and tangential velocity ($V_{\rm  T}$). Stellar ages were estimated with the CANNON. 

We fit the velocity distributions with two different models, namely the GMM and the Gauss-Hermite polynomial. In the bulge region, the Gauss-Hermite coefficients $h_{3}$ are positively correlated with the mean velocity $\olsi{V}$, which may be due to the Galactic bar structure. At $|l|>10\degree$ (the disk dominated region), $h_{3}$ is anti-correlated with $\olsi{V}$, consistent with theoretical expectations.

Fitting the $V_{\rm GSR}$ profiles with a 2-component GMM we find that most of the secondary components in the bulge region are not dynamically cold (with a threashold of $\sigma<40\kms$). Although some key fields within $-10\degree < l < -5\degree$ lack the coverage, we still find symmetric longitudinal trends in mean velocity ($\mu$) and standard deviation ($\sigma$) distributions of the secondary component, which imply a bar-like kinematic. In the bulge region, the coldest HV peak show up at $|l|\sim6\degree$. If we free the number of Gaussians in GMM fitting, $n=3-4$ are usually required, implying a complex velocity distribution in the MW bulge. With the additional tangential motion information from $Gaia$, we find that the HV stars show similar patterns in the radial-tangential velocity distribution $(V_{\rm  R}-V_{\rm  T})$, regardless of the existence of a distinct cold HV peak. 

Moreover, without specifically selecting certain kinds of resonant orbits, a simple MW bar model could reproduce well the observed $V_{\rm  R}-V_{\rm  T}$ (or $V_{\rm  GSR}-\mu_{l}$) distributions. The chemical abundances and the age inferred from the ASPCAP and the CANNON implies that, the HV stars in the bulge/bar region $(R < 4\kpc)$ are generally as old as the other stars in the bar region.

Unfortunately, currently no single model can well explain all the observational results. The models mainly fall into two categories, i.e., the $x_{1}$-orbit scenario and the $x_{2}$-orbit scenario:

(1) The $x_{1}$-orbit scenario: \citet{Aumer2015} argued that the HV peak is contributed by young stars captured on $x_{1}$-orbits by the bar potential, since the selection function of APOGEE might be more sensitive to young stars. In our study, we found that the APOGEE-2 HV stars do show bar-like kinematics. $x_1$ orbits are the back-bone of the bar, and therefore they likely contribute to the observed HV feature, but in observations we also found that these HV stars are not young (older than 8 Gyrs). \cite{McGough2020} presented a new model for galactic bars, in which propeller orbits (a ``distant relative'' of $x_{1}$ orbits) plays a dominant role in the orbital structure. They suggested that the propeller orbits, which are not necessarily young, may be responsible for the observed HV peaks, since their proper motions match the observations well. Fig.~\ref{fig:7} shows that without selecting certain kinds of orbits, a barred model could also well reproduce the observed $V_{\rm GSR}-\mu_{l}$ distributions. This indicates that other orbital families, different from the propeller orbits, can also form similar HV features. Thus, the propeller orbits are sufficient but not necessary to explain the HV features. 

(2) The $x_2$-orbit scenario: For the kpc-scale nuclear disk model composed by $x_2$ orbits \citep{Debattista2015, Debattista2018}, there are three predictions: the HV peaks have (a) smaller $|\mu|$ and (b) smaller $|\sigma|$ at negative longitudes, in comparison to the corresponding positive longitudes with the same $|l|$; (c) the nuclear disk must form later than the stars in the main component, hinting for a possible age difference with the HV peak stars, which are younger than the main component. According to the observational data, the fields at $l = \pm10^\circ$ are inconsistent with all these predictions. In addition, the longitudinal symmetric distributions of $\mu, \sigma$, and $h_{3}$ of the secondary component shown in Figures \ref{fig:3}-\ref{fig:5} are inconsistent with the $x_{2}$-orbit predictions. More observations and theoretical efforts are needed to fully understand the origin of the HV peaks/shoulders in the bulge/bar region.

\acknowledgments
We thank the referee for the constructive and valuable comments that helped to improve this paper. The research presented here is partially supported by the National Key R\&D Program of China under grant No. 2018YFA0404501; by the National Natural Science Foundation of China under grant Nos. 11773052, 11761131016, 11333003; by the ``111'' Project of the Ministry of Education under grant No. B20019, and by the MOE Key Lab for Particle Physics, Astrophysics and Cosmology. This work made use of the Gravity Supercomputer at the Department of Astronomy, Shanghai Jiao Tong University, and the facilities of the Center for High Performance Computing at Shanghai Astronomical Observatory.

J.G.F-T is supported by FONDECYT No. 3180210 and Becas Iberoamérica Investigador 2019, Banco Santander Chile.

Funding for the Sloan Digital Sky Survey IV has been provided by the Alfred P. Sloan Foundation, the U.S. Department of Energy Office of Science, and the Participating Institutions. SDSS-IV acknowledges
support and resources from the Center for High-Performance Computing at
the University of Utah. The SDSS web site is www.sdss.org.

SDSS-IV is managed by the Astrophysical Research Consortium for the 
Participating Institutions of the SDSS Collaboration including the 
Brazilian Participation Group, the Carnegie Institution for Science, 
Carnegie Mellon University, the Chilean Participation Group, the French Participation Group, Harvard-Smithsonian Center for Astrophysics, 
Instituto de Astrof\'isica de Canarias, The Johns Hopkins University, Kavli Institute for the Physics and Mathematics of the Universe (IPMU) / 
University of Tokyo, the Korean Participation Group, Lawrence Berkeley National Laboratory, 
Leibniz Institut f\"ur Astrophysik Potsdam (AIP),  
Max-Planck-Institut f\"ur Astronomie (MPIA Heidelberg), 
Max-Planck-Institut f\"ur Astrophysik (MPA Garching), 
Max-Planck-Institut f\"ur Extraterrestrische Physik (MPE), 
National Astronomical Observatories of China, New Mexico State University, 
New York University, University of Notre Dame, 
Observat\'ario Nacional / MCTI, The Ohio State University, 
Pennsylvania State University, Shanghai Astronomical Observatory, 
United Kingdom Participation Group,
Universidad Nacional Aut\'onoma de M\'exico, University of Arizona, 
University of Colorado Boulder, University of Oxford, University of Portsmouth, 
University of Utah, University of Virginia, University of Washington, University of Wisconsin, 
Vanderbilt University, and Yale University.

This work has made use of data from the European Space Agency (ESA) mission
{\it Gaia} (\url{https://www.cosmos.esa.int/gaia}), processed by the {\it Gaia}
Data Processing and Analysis Consortium (DPAC,
\url{https://www.cosmos.esa.int/web/gaia/dpac/consortium}). Funding for the DPAC
has been provided by national institutions, in particular the institutions
participating in the {\it Gaia} Multilateral Agreement.

\bibliographystyle{apj}
\bibliography{mybib}

\appendix
Fig. \ref{fig:ap1} shows the comparison between the test data from APOKASC and the predicted values from the CANNON.

\renewcommand{\thefigure}{A\arabic{figure}}
\setcounter{figure}{0}

\begin{figure}[!t]
\centering

\includegraphics[width=\columnwidth]{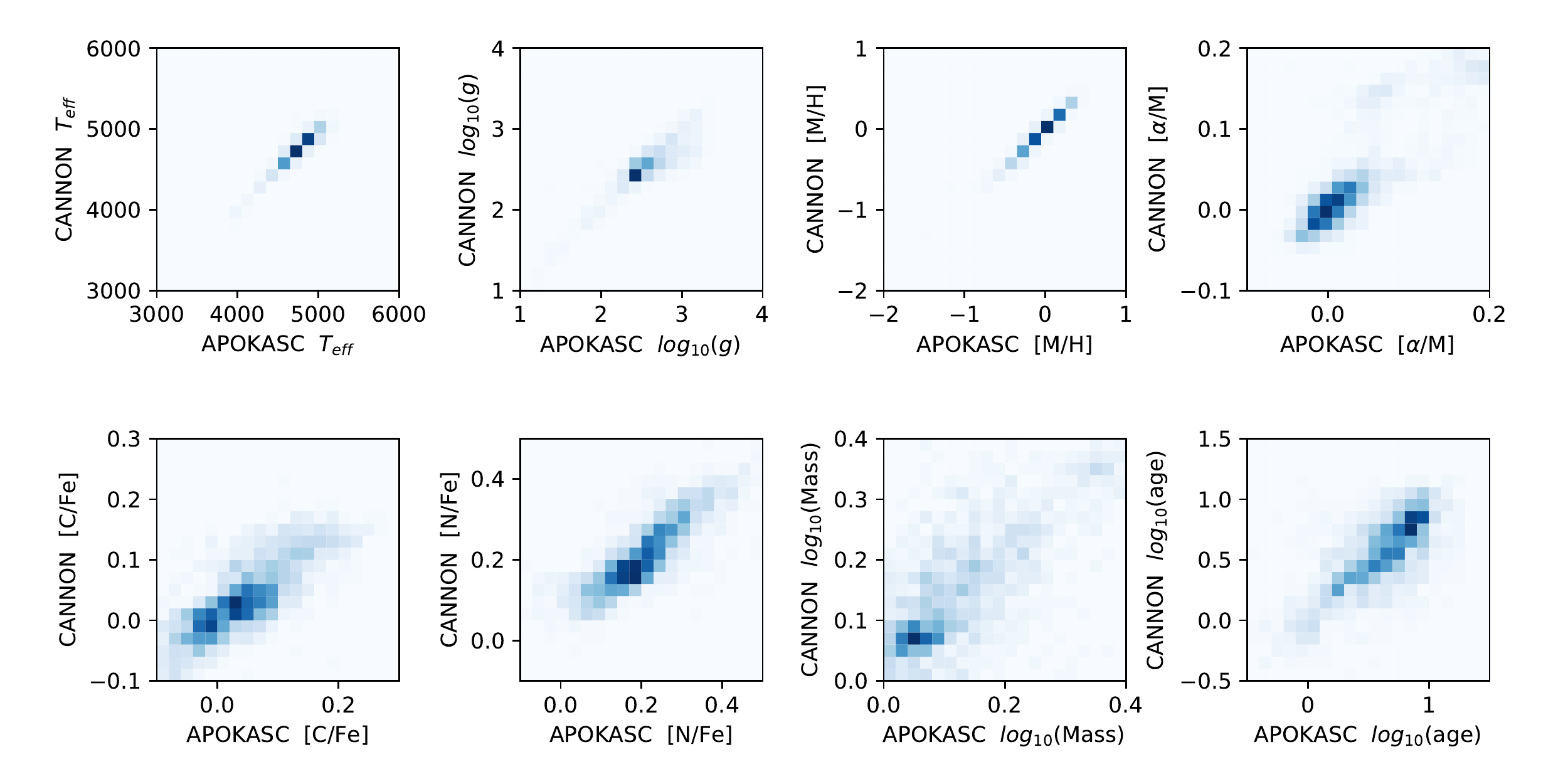}
\centering\caption{CANNON result. x: test data from APOKASC; y: predict values from the CANNON.}
\label{fig:ap1}
\end{figure}



\end{document}